\documentclass[12pt,twoside]{article} 
\usepackage[T1]{fontenc} % font

\usepackage{amsthm, amsmath, amssymb, mathrsfs, upgreek, enumerate, mathtools,
  comment, natbib, subfigure, graphicx, mathabx, booktabs, threeparttable,
tikz, relsize, exscale, epstopdf, scrextend, multirow, xr-hyper, pdfpages, bm}
\usepackage[headheight=110pt,margin=1.25in]{geometry}
\usepackage[section]{placeins}

\usepackage[toc,page,header]{appendix}

\usepackage[labelsep=period]{caption}
\usepackage{fancyhdr, setspace} % headers and footers
\usepackage[pagebackref=false]{hyperref} % link colors
\hypersetup{
    colorlinks=true,
    citecolor=blue,
    filecolor=black,
    linkcolor=black,
    urlcolor=blue,
    bookmarksopen=true,
    pdfstartview=FitH
}

\newtheorem{theorem}{Theorem}
\newtheorem{assump}{Assumption}
\newtheorem{assump*}{Assumption}[section]
\newtheorem{proposition}{Proposition}

\theoremstyle{definition}
\newtheorem{definition}{Definition}

\newtheorem{remark}{Remark}

\def\equationautorefname~#1\null{(#1)\null} % for autoref to work as eqref

\def\Snospace~{\S{}}
\makeatletter
\def\thm@space@setup{
  \thm@preskip=15pt \thm@postskip=15pt % controls spacing before and
}
\makeatother

\def\indep{\perp\!\!\!\perp}

\newcommand{\cov}{\text{Cov}}
\newcommand{\var}{\text{Var}}
\newcommand{\E}{{\bf E}}
\newcommand{\R}{\mathbb{R}}
\newcommand{\N}{\mathbb{N}}
\newcommand{\prob}{\mathbf{P}}
\newcommand{\ind}{\bm{1}}
\newcommand{\plimarrow}{\stackrel{p}\longrightarrow}
\newcommand{\dlimarrow}{\stackrel{d}\longrightarrow}

\newcommand*{\medcup}{\mathbin{\scalebox{1.5}{\ensuremath{\cup}}}}

\providecommand{\abs}[1]{\lvert#1\rvert} 
\providecommand{\norm}[1]{\lVert#1\rVert}

\renewcommand{\qed}{\hfill \mbox{\raggedright \rule{0.08in}{0.08in}}} % black QED box
\renewenvironment{proof}[1][\proofname]{{\noindent\sc#1. }}{\qed\vspace{15pt}} % "proof" small caps

\title{Causal Inference Under Approximate Neighborhood Interference\bf\sc \thanks{This research is supported by National Science Foundation grant SES-1755100. I thank Guido Imbens for suggestions that led to a simpler formulation of \autoref{aani}. I also thank Dean Eckles, Fredrik S\"{a}vje, and seminar audiences at Harvard, MIT, Northwestern, Rutgers, Stanford, UCL, and the 2019 SAMSI Causal Inference Workshop for useful feedback.}}

\author{Michael P.\ Leung\thanks{Department of Economics, University of Southern California. E-mail: leungm@ucsc.edu.}}

\begin{document}
\maketitle

\begin{abstract}

  {\sc Abstract.} This paper studies causal inference in randomized experiments under network interference. Commonly used models of interference posit that treatments assigned to alters beyond a certain network distance from the ego have no effect on the ego's response. However, this assumption is violated in common models of social interactions. We propose a substantially weaker model of ``approximate neighborhood interference'' (ANI) under which treatments assigned to alters further from the ego have a smaller, but potentially nonzero, effect on the ego's response. We formally verify that ANI holds for well-known models of social interactions. Under ANI, restrictions on the network topology, and asymptotics under which the network size increases, we prove that standard inverse-probability weighting estimators consistently estimate useful exposure effects and are approximately normal. For inference, we consider a network HAC variance estimator. Under a finite population model, we show that the estimator is biased but that the bias can be interpreted as the variance of unit-level exposure effects. This generalizes Neyman's well-known result on conservative variance estimation to settings with interference.
  
  \vspace{15pt}


 
\end{abstract}

\newpage

%----------------------------------------------------------------------
\section{Introduction}\label{sintro}
%---------------------------------------------------------------------- 
\onehalfspacing

Randomized experiments in settings with network interference have seen increasing use in economics and the social sciences.\footnote{E.g.\ \cite{bandiera2009social}, \cite{bond201261}, \cite{bursztyn_understanding_2014}, \cite{miguel2004worms}, \cite{paluck2016changing}.} This paper develops methods for causal inference in experiments under relatively weak restrictions on interference. We consider a finite population of $n$ units connected through a network $\bm{A}$. Let $Y_i(\bm{d})$ denote the potential outcome of unit $i$ under the counterfactual that the network is assigned treatment vector $\bm{d} = (d_i)_{i=1}^n \in \{0,1\}^n$. The dependence of $Y_i(\bm{d})$ on the entire vector of assignments allows for interference or ``spillovers,'' in contrast to the standard potential outcomes model. Interest centers on ``exposure effects,'' defined below, which summarize how outcomes change in response to manipulations of $\bm{d}$.

The main inferential challenge is that, with a single network, the econometrician observes only one realization of the treatment assignment vector $\bm{D} = (D_i)_{i=1}^n$, where $D_i \in \{0,1\}$ denotes unit $i$'s realized assignment. Identification of exposure effects is therefore impossible without restrictions on the manner in which $Y_i(\cdot)$ varies with $\bm{d}$. The predominant approach in the literature is to assume interference operates through a low-dimensional vector of sufficient statistics.\footnote{E.g.\ \cite{aronow_estimating_2017}, \cite{basse2019randomization}, \cite{forastiere2020identification}, \cite{manski2013identification}, \cite{toulis2013estimation}.} That is, $Y_i(\bm{D})$ is only a function of $\bm{D}$ through a vector-valued {\em exposure mapping} 
\begin{equation*}
  T_i \equiv T(i,\bm{D},\bm{A})
\end{equation*}

\noindent whose dimension is fixed with respect to $n$, unlike that of $\bm{D}$.

For example, \cite{cai2015social} run an experiment to study the effect of providing information on the benefits of weather insurance on farmers' take-up of insurance. The authors are interested in spillover effects since a farmer who obtains the information may disseminate it to her social contacts. They estimate linear versions of the model
\begin{equation}
  Y_i(\bm{D}) = \tilde Y_i(T_i), \quad\text{where}\quad T_i = \left( D_i, \frac{\sum_j A_{ij} D_j}{\sum_j A_{ij}} \right) \label{aseg}
\end{equation}

\noindent and $A_{ij}$ is an indicator for whether farmers $i$ and $j$ are friends. Here the exposure mapping $T_i$ is two-dimensional. Variation in its first component identifies the direct effect of the intervention (being offered to attend an information session on weather insurance), while variation in the second identifies a spillover effect.

Exposure mappings, if correctly specified, substantially reduce the dimensionality of the model since we can reparameterize potential outcomes as
\begin{equation}
  Y_i(\bm{d}) = \tilde Y_i(t) \quad\text{for}\quad t \in \mathcal{T}, \label{reparam}
\end{equation}

\noindent where $\mathcal{T}$ is the range of $T(\cdot)$. Interest then centers on ``exposure effects''
\begin{equation}
  \frac{1}{n} \sum_{i=1}^n \big(\tilde Y_i(t) - \tilde Y_i(t')\big) \label{parsimony}
\end{equation}

\noindent for $t,t' \in \mathcal{T}$, which measure the average change in potential outcomes in response to counterfactual manipulations of the exposure mapping. 

The majority of the literature studies \eqref{reparam} under the assumption that $T_i$ only depends on treatments assigned to the {\em $K$-neighborhood} of $i$ for some small $K$.\footnote{This is the set of units whose network distance from $i$ is at most $K$, formally defined in \autoref{ssetup}.} However, this imposes strong structural restrictions on the underlying outcome process that are incompatible with a variety of models of social interactions studied in the networks literature \citep{guilbeault2018complex,jackson2010,manski1993identification}. In these models, interference can arise from units outside of the ego's $K$-neighborhood, for any $K$. This is the case for models with endogenous peer effects, where outcomes are functions of the outcomes of neighbors \citep{eckles2017design}. Another example is the \cite{cai2015social} setting, where under a simple diffusion model, information obtained by treated units can eventually diffuse to distant alters, which violates \eqref{aseg}.

Important recent work studies {\em misspecified} exposure mappings \citep{chin2018central,savje2017average,savje2021causal}. The insight of this literature is that standard estimators for \eqref{parsimony} unbiasedly estimate meaningful exposure effects even without imposing \eqref{reparam} to restrict interference, which indicates a certain robustness of the estimator to more general patterns of interference. However, under what conditions {\em inference} can be made similarly robust is a more challenging question. Without \eqref{reparam}, potential outcomes can depend arbitrarily on the entire assignment vector $\bm{D}$, which makes large-sample inference impossible. These papers accordingly propose a variety of high-level conditions weaker than \eqref{reparam} that implicitly restrict interference in order to obtain large-sample results. Unfortunately, the connection between these conditions and the literature on social interactions remains unclear. It is an open question whether models in the networks literature violating \eqref{reparam} (e.g.\ models with endogenous peer effects or other diffusion models) can satisfy these high-level conditions.

%-----

\paragraph{Our Contribution.} We study inference on exposure effects with misspecified exposure mappings under a new restriction on interference. We propose a model of {\em approximate neighborhood interference} (ANI), which allows treatments assigned to units further from the ego to have potentially nonzero, but smaller, effects on the ego's response. Unlike the existing literature, we formally verify ANI in well-known models of social interactions. We also show that, under ANI, the data satisfies {\em $\psi$-dependence}, a recently proposed notion of weak network dependence. This enables us to apply limit theorems due to \cite{kojevnikov2021limit} to establish that, under restrictions on the network topology, standard inverse probability weighting (IPW) estimators are consistent for certain exposure effects and asymptotically normal. 

For inference, we consider a network HAC (heteroskedasticity and autocorrelation consistent) variance estimator and characterize its asymptotic bias under a finite population model. We show the bias can be interpreted as the variance of unit-level exposure effects, which is fundamentally unidentified even under no interference. This generalizes the well-known result on conservativeness of the standard variance estimator of the difference-in-means estimate under no interference \citep[e.g.][Ch.\ 6.4.2]{imbens_causal_2015} to settings with dependence due to interference. 

Finally, we propose a novel bandwidth for the network HAC estimator based on the average path length of the observed network. For a given bandwidth rule, the HAC estimator has different rates of convergence depending on whether the average $K$-neighborhood size in $\bm{A}$ grows exponentially or polynomially with $K$. The utility of the average path length is that its magnitude adapts to the neighborhood growth rate to better trade-off bias and variance. 

\cite{kojevnikov2021limit} and \cite{kojevnikov2021bootstrap} respectively provide consistency results for network HAC and bootstrap variance estimators for $\psi$-dependent network data. Their results pertain to settings in which the data is mean-homogeneous, which often holds in superpopulation models. We extend their results to settings with mean-heterogeneous data, as is the case in finite population models.

\cite{choi2017estimation} and \cite{choi2018using} study causal inference without imposing an exposure mapping model. These papers focus on different estimands than ours and assume treatment responses satisfy a monotonicity condition, which we do not require. There is also work on testing for interference, which can be used to test for correct specification of exposure models \citep[e.g.][]{athey2018exact}.

Several papers in econometrics study causal inference under interference \citep{baird2018optimal,he2021measuring,lazzati2015treatment,leung2017treatment,vazquez2017identification,viviano2019policy}. The second paper studies a dynamic setting, whereas the others focus on a static setting like ours but under correctly specified exposure mappings. Many assume the special case of {\em stratified interference} under which the data consists of many clusters and interference only operates within clusters. We instead consider a single large cluster with known network structure.

The next section states our basic assumptions. In \autoref{sani}, we present our model of interference (ANI) and large-sample results. We discuss variance estimation in \autoref{svar}. In \autoref{snumill}, we illustrate the performance of our methods in an empirical application and simulation study. Finally, \autoref{sconclude} concludes. All proofs are given in \autoref{sproofs}.

%----------------------------------------------------------------------
\section{Setup}\label{ssetup}
%----------------------------------------------------------------------

We consider a finite population model in which the only source of randomness is {\em design uncertainty} \citep{abadie2019sampling,imbens2019causal}. That is, $\bm{D}$ is the only random quantity. In the special case of no interference, this corresponds to the well-known Neyman causal model. The setup can be viewed as conditioning on the network and potential outcomes, which allows for arbitrary dependence between the two. This allows links to form at higher rates between units with similar unobservables, which corresponds to unobserved homophily, a well-known hindrance to identifying social interactions \citep{shalizi2011homophily}.

Let $\mathcal{N}_n = \{1, \dots, n\}$ denote the set of units. We assume $\bm{A}$ is an undirected and unweighted network with no self-links, represented as an adjacency matrix with $ij$th entry $A_{ij} \in \{0,1\}$ denoting a potential link between units $i$ and $j$. Let $\mathcal{A}_n$ denote the set of such networks on $n$ units. We assume the components of $\bm{D}$ are independent across units but not necessarily identically distributed, which allows for assignment based on unit covariates and network position. For example, treatment may be assigned ``optimally'' according to these characteristics \citep[e.g.][]{viviano2019policy}, or randomization may be stratified, as in the empirical application in \autoref{sapp}.

Let $\mathcal{T} \subseteq \R^{d_T}$ be a discrete set. For any $n\in\mathbb{N}$, an {\em exposure mapping} is a function $T \equiv T_n\colon \mathcal{N}_n \times \{0,1\}^n \times \mathcal{A}_n \rightarrow \mathcal{T}$. Most of the literature assumes $T(\cdot)$ is {\em correctly specified} in the sense that \eqref{reparam} holds, which enables a simple definition of exposure effects \eqref{parsimony}. We instead follow the literature on misspecified exposure mappings and employ $T(\cdot)$ only to define useful estimands that summarize treatment and spillover effects but not to restrict the true interference structure. This is a reasonable solution to the task of parsimoniously summarizing the causal effect of a high-dimensional vector $\bm{D}$ on potential outcomes.

For any $n\in\mathbb{N}$ and $i\in\mathcal{N}_n$, a potential outcome $Y_i(\cdot)$ is a mapping from $\{0,1\}^n$ to $\R$. Define the {\em unit-level exposure effect}
\begin{equation*}
  \tau_i(t,t') = \mu_i(t) - \mu_i(t'), \quad\text{where}\quad \mu_i(t) = \sum_{\bm{d} \in \{0,1\}^n} Y_i(\bm{d}) \prob(\bm{D}=\bm{d} \mid T_i = t)
\end{equation*}

\noindent and $t,t' \in \mathcal{T}$. This is the difference in unit $i$'s expected response under two different values of the exposure mapping. The estimand of interest is the average effect
\begin{equation}
  \tau(t,t') = \mu(t) - \mu(t'), \quad\text{where}\quad \mu(t) = \frac{1}{n} \sum_{i=1}^n \mu_i(t), \label{aee}
\end{equation}

\noindent also studied by \cite{savje2021causal} and analogous to estimands proposed by \cite{hudgens2008toward} but generalized to allow for an incomplete network. We refer to these references for more detailed discussion of interpretation, but the idea is analogous to \eqref{parsimony}, which is to compare the average outcomes of units under two different values of the exposure mapping. 

Recall $T_i \equiv T(i, \bm{D}, \bm{A})$, and define the generalized propensity score \citep{imbens2000role}
\begin{equation*}
  \pi_i(t) = \E[\ind_i(t)], \quad\text{where}\quad \ind_i(t) = \ind\{T_i=t\}.\footnote{Computation of $\pi_i(t)$ depends on the exposure mapping and design. In \autoref{sapp}, where treatments are block randomized, it can be easily computed in closed form using the hypergeometric distribution.}
\end{equation*}

\noindent We estimate $\tau(t,t')$ using the standard IPW estimator, which is unbiased:
\begin{equation*}
  \hat\tau(t,t') = \hat\mu(t) - \hat\mu(t'), \quad\text{where}\quad \hat\mu(t) = \frac{1}{n} \sum_{i=1}^n Y_i \frac{\ind_i(t)}{\pi_i(t)}.\footnote{The ``H\'{a}jek estimator'' replaces $\hat\mu(t)$ with $\sum_{i=1}^n Y_i \frac{\ind_i(t)}{\pi_i(t)} / \sum_{i=1}^n \frac{\ind_i(t)}{\pi_i(t)}$, which improves efficiency at the cost of finite-sample bias \citep[][\S 7.2]{hirano2003efficient,aronow_estimating_2017}. It is straightforward to extend our results to this estimator by deriving the usual asymptotically linear representation.}
\end{equation*}

Most of the literature focuses on {\em $K$-neighborhood exposure mappings}, requiring $T(\cdot)$ to only be a function of $\bm{d}$ and $\bm{A}$ through $i$'s $K$-{\em neighborhood}, denoted
\begin{equation*}
  \mathcal{N}_{\bm{A}}(i,K) = \{j\in\mathcal{N}_n\colon \ell_{\bm{A}}(i,j) \leq K\},
\end{equation*}

\noindent where $\ell_{\bm{A}}(i,j)$ is the {\em path distance} between $i,j$.\footnote{A {\em path} between $i,j$ is a sequence of links $A_{k_1k_2}, A_{k_2k_3}, \dots, A_{k_{m-1}k_m}=1$ such that $k_1=i$, $k_m=j$, and $k_a \neq k_b$ for all $a,b \in \{1, \dots, m\}$. The {\em length} of this path is $m-1$. The path distance between $i,j$ is the length of the shortest path between them, defined as $\infty$ if $i\neq j$ and no path exists and defined as 0 if $i=j$.} (Note that $i \in \mathcal{N}_{\bm{A}}(i,K)$.) To formalize this requirement, define $\bm{d}_{\mathcal{N}_{\bm{A}}(i,K)} = (d_j\colon j \in \mathcal{N}_{\bm{A}}(i,K))$ and $\bm{A}_{\mathcal{N}_{\bm{A}}(i,K)} = (A_{kl}\colon k,l \in \mathcal{N}_{\bm{A}}(i,K))$, respectively the subvector of $\bm{d}$ and subnetwork of $\bm{A}$ on $\mathcal{N}_{\bm{A}}(i,K)$. 

\begin{assump}[Exposure Mappings]\label{aKexposure}
  There exists $K \in \N$ such that, for any $n\in\N$ and $i\in\mathcal{N}_n$, $T(i,\bm{d},\bm{A}) = T(i,\bm{d}',\bm{A}')$ for all $\bm{d},\bm{d}' \in \{0,1\}^n$ and $\bm{A},\bm{A}'\in\mathcal{A}_n$ such that $\mathcal{N}_{\bm{A}}(i,K) = \mathcal{N}_{\bm{A}'}(i,K)$, $\bm{A}_{\mathcal{N}_{\bm{A}}(i,K)} = \bm{A}_{\mathcal{N}_{\bm{A}'}(i,K)}'$, and $\bm{d}_{\mathcal{N}_{\bm{A}}(i,K)} = \bm{d}_{\mathcal{N}_{\bm{A}'}(i,K)}'$.
\end{assump}

\noindent This is a weak restriction on $T(\cdot)$ satisfied by most exposure mappings of interest in the literature. See for example \eqref{aseg} or the estimands in \autoref{snumill}, where $K=1$.

The next two assumptions impose uniform boundedness on potential outcomes and the generalized propensity score, which will involve bounding these quantities over all $n$. For this to be well-defined, we embed the observed network $\bm{A}$ in a sequence of networks $\{\bm{A}_m\}_{m\in\mathbb{N}}$ where $\bm{A}_m$ is a network on $m$ nodes and $\bm{A}_n = \bm{A}$. We will be imposing conditions on this sequence, which amount to restrictions on the topology of the observed network for large $n$; see \autoref{stheory} for further discussion.

\begin{assump}[Overlap]\label{aoverlap}
  $\pi_i(t) \in [\underline{\pi},\overline{\pi}] \subset (0,1)$ $\forall\,n\in\mathbb{N}$, $i\in\mathcal{N}_n$, $t\in\mathcal{T}$.
\end{assump}

\noindent This requires the generalized propensity score to be uniformly bounded away from 0 and 1 over the sequence of networks. While overlap is standard, it can be restrictive. In particular, it restricts the exposure mapping, network sequence, distribution of treatments, and population. For instance, if $T_i = \mathbf{1}\{\sum_j A_{ij}D_j > 0\}$ and treatments are i.i.d., then overlap holds for $\pi_i(1)$ if treatment assignment is nontrivial and degrees $\sum_j A_{ij}$ are uniformly bounded over $i,n$. However, if $i$'s degree is large, then $\pi_i(1)$ will be close to one since it is highly likely to have at least one of many neighbors treated. At the cost of changing the estimand, overlap can be restored if we instead randomize treatment only to a small subset of ``eligible'' units and restrict the population to units with an eligible neighbor. Then even with a large degree, the chance of having an {\em eligible} neighbor treated can be far from 0 and 1. This is the design in \autoref{sapp}. 

\begin{assump}[Bounded Outcomes]\label{aYbound}
  $\abs{Y_i(\bm{d})} < \overline{Y} < \infty$ $\forall\, n\in\mathbb{N}$, $i\in\mathcal{N}_n$, $\bm{d}\in\{0,1\}^n$.
\end{assump}

\noindent This assumption is standard and can be generalized to uniformly bounded moments.

%----------------------------------------------------------------------
\section{Approximate Neighborhood Interference}\label{sani}
%---------------------------------------------------------------------- 

We next present our model of interference. For any $\bm{d} \in \{0,1\}^n$, partition $\bm{d} =  (\bm{d}_{\mathcal{N}_{\bm{A}}(i,s)}, \bm{d}_{-\mathcal{N}_{\bm{A}}(i,s)})$, so that $\bm{d}_{-\mathcal{N}_{\bm{A}}(i,s)} = (d_j\colon j \in \mathcal{N}_n\backslash\mathcal{N}_{\bm{A}}(i,s))$. Let $\bm{D}'$ be an independent copy of $\bm{D}$. Define $\bm{D}^{(i,s)} = (\bm{D}_{\mathcal{N}_{\bm{A}}(i,s)}, \bm{D}'_{-\mathcal{N}_{\bm{A}}(i,s)})$, obtained by concatenating the subvector of $\bm{D}$ on $\mathcal{N}_{\bm{A}}(i,s)$ and that of $\bm{D}'$ on $\mathcal{N}_n\backslash\mathcal{N}_{\bm{A}}(i,s)$. Finally, let
\begin{equation*}
  \theta_{n,s} \equiv \max_{i\in\mathcal{N}_n} \E[\abs{Y_i(\bm{D}) - Y_i(\bm{D}^{(i,s)})}].
\end{equation*}

\noindent This measures interference from ``distant'' alters, those more than distance $s$ away from the ego. It is the largest expected perturbation of any unit's potential outcome due to redrawing the treatment assignments of distant alters.

\begin{assump}[ANI]\label{aani}
  $\sup_n \theta_{n,s} \rightarrow 0$ as $s\rightarrow\infty$.
\end{assump}

\noindent This requires interference from distant alters to be negligible for large distances. In the special case of correct specification \eqref{reparam}, there is {\em no} interference from units outside $i$'s $K$-neighborhood, so $Y_i(\bm{D}) - Y_i(\bm{D}^{(i,s)})=0$ for all $s\geq K$ and any $i$. In contrast, ANI allows this difference to be nonzero for all $s$ but requires that it decays with $s$. This means interference from alters beyond a unit's $s$-neighborhood becomes increasingly negligible as we expand the radius $s$. Hence, ANI says that a unit's response is primarily, but not entirely, determined by the assignments of alters close to it.\footnote{\autoref{aani} has some similarities with Assumption 6 of \cite{chin2018central} in bounding the effect of manipulations of treatment assignments of distant units. For a CLT, we do not need a high-level condition analogous to his Assumption 5, which requires correlations between observed outcomes $\{Y_i\}_{i=1}^n$ to be sufficiently weak.}

ANI restricts the network, potential outcomes, and distribution of treatments. The first two are quite evident since the existence of interference is determined by $\bm{A}$ as well as the potential outcomes $Y_i(\cdot)$. For instance, if $\bm{A}$ has no links for any $n$, then ANI holds because for $s>0$, $Y_i(\bm{D}) - Y_i(\bm{D}^{(i,s)})=0$ for any potential outcome model, as no interference is possible without neighbors. On the other hand, for some $Y_i(\cdot)$, ANI can potentially hold for any $\bm{A}$. In the SUTVA case where $Y_i(\bm{d}) = \tilde Y_i(d_i)$ for all $\bm{d}$, this is clearly the case since for any $\bm{A}$ and $s>0$, $Y_i(\bm{D}) - Y_i(\bm{D}^{(i,s)})=0$.\footnote{As pointed out by a referee, since $\theta_{n,s}$ is defined with respect to the design, it may be possible to choose the design to ensure that ANI holds for a given network and potential outcome model. See for example \autoref{thresholdmodel} below, where a sufficient condition for ANI depends on a parameter $\varphi_j$ that is a function of the distribution of $D_j$.}

%---------------------------------------
\subsection{Social Interactions Models}\label{splaus}
%---------------------------------------

We next verify ANI for two well-known models of social interactions. Our results yield uniform bounds on $\theta_{n,s}$ that decay exponentially with $s$ under restrictions on the strength of social interactions.

\bigskip

\noindent {\bf Linear-in-Means Model.} Consider a network version of the \cite{manski1993identification} model
\begin{equation}
  Y_i = \alpha + \beta \frac{\sum_j A_{ij}Y_j}{\sum_j A_{ij}} + D_i\gamma + \varepsilon_i, \label{LIMM}
\end{equation}

\noindent where the unobserved heterogeneity $\{\varepsilon_i\}_{i=1}^n$ is uniformly bounded (to impose \autoref{aYbound}) and nonrandom. As usual, to ensure the model is coherent, we assume
\begin{equation}
  |\beta|<1.
  \label{LIMweak}
\end{equation}

\noindent The model defines potential outcomes $Y_i(\bm{D})$ through its reduced form
\begin{equation*}
  \bm{Y} = \frac{\alpha}{1-\beta} \ind + \bm{D}\gamma + \gamma \beta \sum_{k=0}^\infty \beta^k \tilde{\bm{A}}^{k+1}\bm{D} + \sum_{k=0}^\infty \beta^k \tilde{\bm{A}}^k \bm{\varepsilon}
\end{equation*}

\noindent \citep[e.g.][eq.\ (6), assuming $\bm{A}$ is connected]{bramoulle2009identification}, where $\bm{Y} = (Y_i)_{i=1}^n$, $\bm{\varepsilon}$ is similarly defined, and $\tilde{\bm{A}}$ is the row-normalized version of $\bm{A}$ (divide each row by its sum). The third term roughly says that the impact of treatments assigned to $k$-neighbors is exponentially down-weighted by $\beta^k$. This leads to the following result.

\begin{proposition}\label{linearinmeans}
  If responses are realized according to the linear-in-means model, then there exists $C>0$ such that \autoref{aani} holds with $\theta_{n,s} \leq C \abs{\beta}^s$ for all $n,s$.
\end{proposition}

\paragraph{Complex Contagion.} We next consider a model of ``complex contagion,'' variants of which have been widely studied in the networks literature.\footnote{E.g.\ \cite{granovetter1978threshold}, \cite{guilbeault2018complex}, \cite{jackson2010}, \cite{montanari2010spread}.} Initialize a dynamic discrete-time process at period 0 at some binary response vector $\bm{Y}^0 \in \{0,1\}^n$, which may be a function of treatments $\bm{D}$ and nonrandom unobserved heterogeneity $\bm{\varepsilon}=(\varepsilon_i)_{i=1}^n$. For some $\R$-valued $\phi(\cdot)$, each unit $i$ at period $t$ updates according to
\begin{equation}
  Y_i^t = \ind\left\{ \beta\frac{\sum_j A_{ij} Y_j^{t-1}}{\sum_j A_{ij}} \geq \phi(D_i,\varepsilon_i) \right\} \label{dynamicTSI}
\end{equation}

\noindent to obtain new responses $\bm{Y}^t = (Y_i^t)_{i=1}^n$ from last period's responses $\bm{Y}^{t-1}$. The rule says that $i$ selects 1 over 0 if and only if the fraction of neighbors choosing response 1 in the previous period is large enough relative to the heterogeneous threshold $\phi(D_i,\varepsilon_i)$. The parameter $\beta$ measures the strength of social interactions. 

Because the setup in \autoref{ssetup} is static, we consider running the dynamic process until the first period $T$ such that $\bm{Y}^T = \bm{Y}^{T-1}$. To ensure such a $T$ exists for any $\bm{Y}^0$, we assume $\beta\geq 0$, the common case of strategic complements \citep{milgrom1990rationalizability}.  We then take $\bm{Y}^T$ as the vector of responses $\bm{Y}$ observed in the data, which yields outcomes $(Y_i(\bm{D}))_{i=1}^n$. Hence, this process implicitly defines potential outcomes. 

To verify \autoref{aani}, we need a condition analogous to \eqref{LIMweak}, which will be more complicated to state since the model is nonlinear. Define a weighted directed network $\bm{G}$ on $\mathcal{N}_n$ with $ij$th entry $G_{ij} = A_{ij}\E[\varphi_j]$ for $\varphi_j = \mathbf{1}\{0 < \phi(D_j,\varepsilon_j) \leq \beta\}$. Let
\begin{equation*}
  \rho_n(\bar{s}) = \sup_{s\geq \bar{s}} \,\norm{\bm{G}^s}_\infty^{1/s} \equiv \sup_{s\geq \bar{s}} \bigg( \max_{i\in\mathcal{N}_n} \sum_{j=1}^n (\bm{G}^s)_{ij} \bigg)^{1/s}
\end{equation*}

\noindent for any $\bar{s} > 0$, so $\norm{\cdot}_\infty$ is the matrix norm induced by the vector $\infty$-norm.

\begin{proposition}\label{thresholdmodel}
  Let $\alpha_{n,s}(\bar{s}) = 2(\rho_n(\bar{s})^{s-1}\mathbf{1}\{s-1\geq \bar{s}\} + \mathbf{1}\{s-1 < \bar{s}\})$. Suppose responses are realized according to the complex contagion model. If $\sup_n \rho_n(\bar{s}) < 1$ for some $\bar{s} > 0$, then for this $\bar{s}$, \autoref{aani} holds with $\theta_{n,s} \leq \alpha_{n,s}(\bar{s})$ for all $n,s$.
\end{proposition}

We next discuss the interpretation of $\sup_n \rho_n(\bar{s}) < 1$ in relation to \eqref{LIMweak}. Let $\bm{I}$ be the $n\times n$ identity matrix. For the linear-in-means model to be coherent, we need $\bm{I}-\beta\tilde{\bm{A}}$ to be invertible, which is true provided $\abs{\beta}<1$. Since $\tilde{\bm{A}}$ is row-normalized, this is equivalent to 
\begin{equation}
  \lambda_\text{max}(\beta\tilde{\bm{A}}) < 1,
  \label{rho1}
\end{equation}

\noindent where $\lambda_\text{max}(\cdot)$ is the spectral radius \citep{bramoulle2009identification}. On the other hand, $\norm{\bm{G}^s}_\infty^{1/s} \stackrel{s\rightarrow\infty}\longrightarrow \lambda_\text{max}(\bm{G})$ by Gelfand's formula, so for any $\epsilon>0$, we can choose $\bar{s}$ large enough such that 
\begin{equation}
  \sup_n \lambda_\text{max}(\bm{G}) < 1 - \epsilon \quad\text{implies}\quad \sup_n \rho_n(\bar{s}) < 1
  \label{rho2}
\end{equation}

\noindent \citep{xl2015}. The left-hand side is clearly analogous to \eqref{rho1}. The difference is that, in $\bm{G}$, we weight each potential link $A_{ij}$ by $\E[\varphi_j]$, whereas in $\beta\tilde{\bm{A}}$, the weight is $\beta/\sum_k A_{ik}$. Both weights are monotonically increasing in $\beta$, so both \eqref{rho1} and \eqref{rho2} restrict the strength of social interactions.

%---------------------------------------
\subsection{Weak Dependence}\label{spsiweak}
%---------------------------------------

Define $Z_i = (\ind_i(t) \pi_i(t)^{-1} - \ind_i(t') \pi_i(t')^{-1}) Y_i$, so that $\hat\tau(t,t') = n^{-1}\sum_{i=1}^n Z_i$. For large-sample inference, we would like the data $\{Z_i\}_{i=1}^n$ to be at most weakly dependent. Recall that $\ell_{\bm{A}}(i,j)$ is the path distance between $i,j$ in $\bm{A}$. Since treatments are independent, the indicators are weakly dependent in the sense that $\ind_i(t) \indep \ind_j(t)$ if $\ell_{\bm{A}}(i,j) > 2K$ by \autoref{aKexposure}. At first glance, $\{Y_i\}_{i=1}^n$ seems to be strongly dependent since $Y_i$ is a function of $\bm{D}$ for all $i$. However, ANI requires $Y_i$ to primarily depend on the treatments of nearby units, which suggests that distant units have weakly dependent outcomes. We next formalize this idea.

We first define a notion of weak network dependence due to \cite{kojevnikov2021limit}. For any $H,H' \subseteq \mathcal{N}_n$, define $\ell_{\bm{A}}(H,H') = \min\{\ell_{\bm{A}}(i,j)\colon i \in H, j \in H'\}$. Let $\bm{Z}_H = (Z_i\colon i \in H)$, $\mathcal{L}_d$ be the set of bounded, $\R$-valued, Lipschitz functions on $\R^d$, and
\begin{equation*}
  \mathcal{P}_n(h,h';s) = \left\{ (H,H')\colon H,H' \subseteq \mathcal{N}_n, |H|=h, |H'|=h', \ell_{\bm{A}}(H,H') \geq s \right\}.
\end{equation*}

\begin{definition}\label{dpsidep}
  A triangular array $\{Z_i\}_{i=1}^n$ is {\em $\psi$-dependent} if there exist (a) uniformly bounded constants $\{\tilde\theta_{n,s}\}_{s, n \in \N}$ with $\theta_{n,0}=1$ $\forall n$ such that $\sup_n \tilde\theta_{n,s} \rightarrow 0$ as $s\rightarrow\infty$, and (b) functionals $\{\psi_{h,h'}(\cdot,\cdot)\}_{h,h'\in\N}$ with $\psi_{h,h'}\colon \mathcal{L}_h \times \mathcal{L}_{h'} \rightarrow [0,\infty)$ such that
  \begin{equation}
    \abs{\cov(f(\bm{Z}_H), f'(\bm{Z}_{H'}))} \leq \psi_{h,h'}(f,f') \,\tilde\theta_{n,s} \label{pwg}
  \end{equation}

  \noindent for all $n,h,h' \in \mathbb{N}$; $s>0$; $f \in \mathcal{L}_h$; $f' \in \mathcal{L}_{h'}$; and $(H,H') \in \mathcal{P}_n(h,h';s)$.
\end{definition}

\noindent This extends temporal $\psi$-dependence \citep{doukhan1999new} to network data by using path in place of temporal distance. The concept says two sets of observations $\bm{Z}_H$ and $\bm{Z}_{H'}$ have small covariance if they are sufficiently distant.

Let $i^*(\cdot)$ be the identity function $x \mapsto x$ on $\R$ and $\mathcal{Z} \subseteq \R$ be any compact set such that $Z_i \in \mathcal{Z}$ for any $n\in\mathbb{N}$ and $i\in\mathcal{N}_n$. Such a set exists by Assumptions \ref{aoverlap} and \ref{aYbound}. For any Lipschitz function $f\colon\R^d\rightarrow\R$, let $\text{Lip}(f)$ its Lipschitz constant and $\norm{f}_\infty=\sup_{x\in\mathcal{Z}^d} \abs{f(x)}$. Let $K$ be the constant in \autoref{aKexposure} and $\lfloor s \rfloor$ be $s$ rounded down to the nearest integer.

\begin{theorem}[Weak Dependence]\label{anipsi}
  Under Assumptions \ref{aKexposure}--\ref{aani}, $\{Z_i\}_{i=1}^n$ is $\psi$-dependent in that \eqref{pwg} holds with $\tilde\theta_{n,s} = \theta_{n,\lfloor s/2 \rfloor}\mathbf{1}\{s>2\max\{K,1\}\} + \mathbf{1}\{s\leq 2\max\{K,1\}\}$ for all $n\in\mathbb{N}$ and $s>0$ and
  \begin{equation*}
    \psi_{h,h'}(f,f') = 2 \left(\norm{f}_\infty \norm{f'}_\infty + h \norm{f'}_\infty \text{Lip}(f) + h' \norm{f}_\infty \text{Lip}(f')\right)
  \end{equation*}

  \noindent for either $h,h' \in \N$, $f \in \mathcal{L}_h$, and $f' \in \mathcal{L}_{h'}$, or $h=h'=1$ and $f=f'=i^*$.
\end{theorem}

%---------------------------------------
\subsection{Large-Sample Theory}\label{stheory}
%---------------------------------------

In what follows, define $\tilde\theta_{n,s}$ as in \autoref{anipsi}. Having established that $\{Z_i\}_{i=1}^n$ is $\psi$-dependent, we can apply results due to \cite{kojevnikov2021limit} to show that $\hat\tau(t,t')$ is consistent and asymptotically normal. Their results require $\tilde\theta_{n,s}$ to decay to zero fast enough, and the speed of decay depends on the network topology, in particular the growth rate of $s$-neighborhood sizes. Intuitively, ANI says that $Z_i$ depends primarily on units in $\mathcal{N}_{\bm{A}}(i,s)$. Hence, if the typical size of these neighborhoods grows rapidly with $s$, then weak dependence requires that this be counterbalanced by having the covariances ($\tilde\theta_{n,s}$) decay to zero faster with $s$. 

The next assumptions formalize these ideas. To clarify their interpretation, in \autoref{discpsiweak}, we verify them for networks with polynomial and exponential neighborhood growth rates when $\tilde\theta_{n,s}$ decays exponentially with $s$, as in the examples in \autoref{splaus}. 

\bigskip

\noindent {\bf Limit Sequence.} Our results take $n\rightarrow\infty$ along a sequence of networks $\{\bm{A}_n\}_{n\in\mathbb{N}}$ defined prior to \autoref{aoverlap}. The results hold for any sequence satisfying our assumptions. The design may implicitly depend on $n$, so the distribution of treatments may also change along the sequence, so long as treatments remain independent across units. Additionally, potential outcomes $Y_i(\cdot)$ and the exposure mapping $T(\cdot)$ may vary with $n$ (and must do so due to the dimensions of their arguments), so long as they satisfy Assumptions \ref{aKexposure} and \ref{aYbound}. We emphasize that our results assume $K$ and $\mathcal{T}$ do {\em not} depend on $n$, but all other quantities may vary along the sequence, within the confines of the stated assumptions.

\bigskip

\noindent Let $\mathcal{N}^\partial_{\bm{A}}(i,s) = \{j\in\mathcal{N}_n\colon \ell_{\bm{A}}(i,j)=s\}$ be the {\em $s$-neighborhood boundary} of $i$, the set of units exactly distance $s$ from $i$, and $M_n^\partial(s) = n^{-1} \sum_{i=1}^n \abs{\mathcal{N}^\partial_{\bm{A}}(i,s)}$, its average size.

\begin{assump}[Weak Dependence for LLN]\label{finitevar}
  $\sum_{s=0}^n M_n^\partial(s) \tilde\theta_{n,s} = o(n)$.
\end{assump}

\noindent This corresponds to Assumption 3.2 of \cite{kojevnikov2021limit}. It restricts the network topology through $M_n^\partial(s)$ and the degree of interference through $\tilde\theta_{n,s}$. Since the former grows with $s$, the latter must decay to zero faster for the sum to be $o(n)$. Furthermore, since $\tilde\theta_{n,s}=1$ for $s\leq 2\max\{K,1\}$, this implies $M_n^\partial(1) = o(n)$, which is a restriction on network density that rules out, for instance, $\bm{A}$ being complete.

It is useful to compare this to its analog for $\alpha$-mixing spatial processes. Consider, for example, Assumption 3(b) of \cite{jenish2009central}, which essentially requires $\sum_{s=1}^\infty s^{d-1} \alpha(s) < \infty$, where $d$ is the dimension of the underlying space and $\alpha(s)$ is the $\alpha$-mixing coefficient, which measures dependence between sets of observations at spatial distance $s$ apart. In the spatial setting, the $s$-neighborhood boundary of $i$ is the set of units at any distance $h \in [s,s+1)$ from $i$. By their Lemma A.1(iii), the size of this set is $O(s^{d-1})$. Thus, we have an analogous trade-off between the sizes of spatial $s$-neighborhood boundaries and the rate of decay of the mixing coefficient.

\begin{theorem}[Consistency]\label{lln}
  Under Assumptions \ref{aKexposure}--\ref{finitevar}, $\abs{\hat\tau(t,t') - \tau(t,t')} \plimarrow 0$.
\end{theorem}

\noindent Inspection of the proof shows that we can sharpen the result to $\abs{\hat\tau(t,t') - \tau(t,t')} = O_p(n^{-1/2})$ if we strengthen \autoref{finitevar} to $\sum_{s=0}^n M_n^\partial(s) \tilde\theta_{n,s} = O(1)$.

Asymptotic normality requires a stronger version of \autoref{finitevar}. Let $M_n(s, k) = n^{-1}\sum_{i=1}^n \abs{\mathcal{N}_{\bm{A}}(i, s)}^k$, the $k$th moment of the $s$-neighborhood size, and
\begin{equation*}
  \mathcal{H}_n(s, m) = \left\{ (i,j,k,l) \in \mathcal{N}_n^4\colon k\in \mathcal{N}_{\bm{A}}(i, m), l \in \mathcal{N}_{\bm{A}}(j, m), \ell_{\bm{A}}(\{i,k\}, \{j,l\}) = s \right\}.
\end{equation*}

\noindent This is the set of paired couples $(i,j)$ and $(k,l)$ such that the units within each couple are at most path distance $m$ apart from one another, and the two pairs are exactly path distance $s$ apart. Define $\sigma_n^2 = \var(n^{-1/2}\sum_{i=1}^n Z_i)$.

\begin{assump}[Weak Dependence for CLT]\label{psiweak}
  There exist $\epsilon>0$ and a sequence of positive constants $\{m_n\}_{n\in\N}$ such that $m_n \rightarrow\infty$ and
  \begin{equation}
    \max\left\{ \sigma_n^{-4} \frac{1}{n^2} \sum_{s=0}^n \abs{\mathcal{H}_n(s, m_n)} \tilde\theta_{n,s}^{1-\epsilon},\quad \sigma_n^{-3} n^{-1/2}M_n(m_n, 2),\quad \sigma_n^{-1} n^{3/2} \tilde\theta_{n,m_n}^{1-\epsilon} \right\} \rightarrow 0. \label{3stooges}
  \end{equation}
\end{assump}

\noindent This strengthens \autoref{finitevar} and corresponds to Assumption 3.4 of \cite{kojevnikov2019limit}.\footnote{The publication version of their paper \citep{kojevnikov2021limit} formulates this assumption slightly differently.} Similar to \autoref{finitevar}, the first term in \eqref{3stooges} requires $\tilde\theta_{n,s}$ to decay to zero fast enough relative to $s$-neighborhood sizes. The second term restricts $s$-neighborhood growth rates, while the third requires sufficiently fast decay of $\tilde\theta_{n,s}$. See \autoref{discpsiweak} for a discussion of the plausibility of this assumption.

\begin{theorem}[Asymptotic Normality]\label{clt}
  Under Assumptions \ref{aKexposure}--\ref{aani} and \ref{psiweak},
  \begin{equation*}
    \sigma_n^{-1} \sqrt{n}\big(\hat\tau(t,t') - \tau(t,t')\big) \dlimarrow \mathcal{N}(0,1).
  \end{equation*}
\end{theorem}

%----------------------------------------------------------------------
\section{Variance Estimation}\label{svar}
%---------------------------------------------------------------------- 

For large-sample inference, we consider the variance estimator
\begin{equation}
  \hat\sigma^2 = \frac{1}{n} \sum_{i=1}^n \sum_{j=1}^n (Z_i - \hat\tau(t,t')) (Z_j - \hat\tau(t,t')) \mathbf{1}\{\ell_{\bm{A}}(i,j)\leq b_n\}, 
  \label{bootstrap}
\end{equation}

\noindent where $Z_i$ is defined at the top of \autoref{spsiweak} and $b_n\geq 0$ is a bandwidth parameter discussed below in \eqref{ourb}. When $b_n=0$, this reduces to the sample variance of the $Z_i$'s, which is only a valid estimator under no interference. Choosing $b_n>0$ places nonzero weight on pairs at most $b_n$ apart in the network, which accounts for possible autocorrelation.\footnote{The estimator is simple to compute. First calculate the path distance matrix $(\ell_{\bm{A}}(i,j))_{i,j\in\mathcal{N}_n}$, which can be done very efficiently for sparse networks using Dijkstra's algorithm (e.g.\ the Python function {\tt dijkstra} in the {\tt scipy.sparse.csgraph} module). Then for the matrix $\bm{P} = (\mathbf{1}\{\ell_{\bm{A}}(i,j)\leq b_n\})_{i,j\in\mathcal{N}_n}$ and vector $\tilde{\bm{Z}} = (n^{-1/2} (Z_i - \hat\tau(t,t')))_{i=1}^n$, we have $\hat\sigma^2 = \tilde{\bm{Z}}'\bm{P}\tilde{\bm{Z}}$.} 

Under correctly specified exposure mappings, we can choose $b_n = 2K$, as in \cite{leung2017treatment}, since $\ind_i(t) \indep \ind_j(t)$ if $\ell_{\bm{A}}(i,j) > 2K$. For misspecified exposure mappings, $b_n$ must grow with $n$ at a rate depending on the network topology. In this case, \eqref{bootstrap} corresponds to a HAC estimator familiar from the time series literature but using network distance in place of temporal distance. This estimator has been previously used in practice \citep[e.g.][]{acemoglu2015state}, and its formal properties were first studied by \cite{kojevnikov2021limit} in a superpopulation setting. Building on their work, we characterize its behavior in a finite population model.

\begin{remark}
  \cite{kojevnikov2021limit} consider general kernel functions that include the uniform kernel in \eqref{bootstrap}. In our simulations in \autoref{smc}, \eqref{bootstrap} is always positive semidefinite (PSD), but this is not theoretically guaranteed.  \cite{kojevnikov2021bootstrap} proposes novel weights motivated by network bootstrap procedures, which are guaranteed to be PSD in finite sample, unlike kernel-based weights. While these weights all have the same asymptotic properties, in simulation experiments, we find that uniform weights better control size in small samples than these alternatives that decay with distance.
\end{remark}

%-----

\paragraph{Choice of Bandwidth.} For $\hat\sigma^2$ to have good large-sample properties, we need to restrict the rate at which $\abs{\mathcal{N}_{\bm{A}}(i,b_n)}$ diverges with $n$ (see \autoref{psiweak2}). That is, how fast $b_n$ can diverge depends on how rapidly $K$-neighborhood sizes grow with $K$. In spatial settings, the rate of growth is polynomial in $K$, so $b_n$ is allowed to diverge at a polynomial rate. (A faster rate is better for bias but worse for variance.) However, in network settings, the rate can be exponential. 

Based on the analysis in \autoref{sbandchoice}, we suggest $b_n$ be chosen as follows. Let $\delta(\bm{A}) = n^{-1} \sum_{i,j} A_{ij}$ be the average degree, $\mathcal{L}(\bm{A})$ the {\em average path length} (APL), and
\begin{equation}
  b_n = \lfloor \max\{\tilde b_n, 2K\} \rceil \quad\text{for}\quad \tilde b_n = \left\{ \begin{array}{cc} \frac{1}{2} \mathcal{L}(\bm{A}) & \text{if } \mathcal{L}(\bm{A}) < 2\frac{\log n}{\log \delta(\bm{A})}, \\ \mathcal{L}(\bm{A})^{1/3} & \text{otherwise}, \end{array} \right. \label{ourb}
\end{equation}

\noindent where $\lfloor\cdot\rceil$ means round to the nearest integer.\footnote{The formula assumes $\delta(\bm{A})>1$, which is typical in practice. The APL is the average value of $\ell_{\bm{A}}(i,j)$ over all pairs in the largest component of $\bm{A}$. A {\em component} of a network is a connected subnetwork such that all units in the subnetwork are disconnected from those not in the subnetwork.} To account for correlation in $\{\ind_i(t)\}_{i=1}^n$ discussed in \autoref{spsiweak}, we set $b_n$ at least equal to $2K$. The fractions $1/2$ and $1/3$ are due to \autoref{psiweak2} below (see \autoref{sbandchoice}). We suggest in practice the researcher report results for several bandwidths in a neighborhood of \eqref{ourb}.

The purpose of comparing $\mathcal{L}(\bm{A})$ and $\log n / \log \delta(\bm{A})$ is to determine whether $K$-neighborhood sizes grow approximately exponentially or polynomially with $K$. As discussed in \autoref{sbandchoice}, in the exponential case, the difference between these two statistics typically converges to zero, whereas in the polynomial case, $\mathcal{L}(\bm{A})$ is much larger, having polynomial order. See, for example, the simulations in \autoref{smc}, which show at least a four-fold difference in APL between the two regimes. Thus, \eqref{ourb} selects a bandwidth of logarithmic (polynomial) order when neighborhood growth rates are approximately exponential (polynomial).\footnote{In the exponential case, we need $b_n = O(\log n)$ for $\var(\hat\sigma^2)$ to be small, which \eqref{ourb} accomplishes since $\mathcal{L}(\bm{A}) \approx \log n / \log \delta(\bm{A})$ in this regime. In the polynomial case, $b_n = O(\log n)$ is also valid, but the bias then vanishes at an extremely slow rate \citep[see][proof of Proposition 4.1]{kojevnikov2021limit}. Our choice of $\mathcal{L}(\bm{A})^{1/3}$ substantially improves this rate since $\mathcal{L}(\bm{A})$ is then polynomial in $n$.}

%-----

\paragraph{Bias of $\hat\sigma^2$.} Define
\begin{align*}
  &\hat\sigma^2_* = \frac{1}{n} \sum_{i=1}^n \sum_{j=1}^n (Z_i - \tau_i(t,t')) (Z_j - \tau_j(t,t')) \mathbf{1}\{\ell_{\bm{A}}(i,j)\leq b_n\} \quad\text{and} \nonumber\\ 
  &R_n = \frac{1}{n} \sum_{i=1}^n \sum_{j=1}^n (\tau_i(t,t') - \tau(t,t')) (\tau_j(t,t') - \tau(t,t')) \mathbf{1}\{\ell_{\bm{A}}(i,j)\leq b_n\}.
\end{align*}

\noindent The former is an ``oracle'' version of $\hat\sigma^2$ that replaces $\hat\tau(t,t')$ with $\tau_i(t,t')$, while the latter is a bias term. \autoref{bootvar} below establishes that
\begin{align}
  &\hat\sigma^2 = \hat\sigma^2_* + R_n + o_p(1) \quad\text{and} \label{decomp}\\
  &\abs{\hat\sigma^2_* - \var(\sqrt{n}\hat\tau(t,t'))} \plimarrow 0. \label{bootconsist}
\end{align}

\noindent Equation \eqref{bootconsist} says that the oracle estimator is consistent for the variance, and \eqref{decomp} says that our estimator is biased. The source of bias is mean-heterogeneity: $\hat\tau(t,t')$ is consistent for $\tau(t,t')$ but not $\tau_i(t,t')$, which is heterogeneous across units. 

The bias $R_n$ has the form of a HAC estimate of the variance of the unit-level exposure effects. It is helpful to compare this to the case of no interference, where $T_i = D_i$, $t=1$, and $t'=0$, so that $\tau(t,t')$ is the usual average treatment effect (ATE). Knowing that units are independent, we can choose $b_n=0$, in which case
\begin{equation*}
  R_n = \frac{1}{n} \sum_{i=1}^n \big(\tau_i(1,0) - \tau(1,0)\big)^2.
\end{equation*}

\noindent This is the well-known asymptotic bias of the standard variance estimator for the difference-in-means estimate of the ATE \citep[e.g.][Theorem 6.2]{imbens_causal_2015}. It measures the variance of the unit-level treatment effects and is generally impossible to estimate in the finite population setting, so the variance estimator is conservative. In the special case of homogeneous unit-level treatment effects, meaning $\tau_i(t,t')$ does not vary with $i$, the bias is zero, a property also shared by our $R_n$. Thus, \eqref{decomp} generalizes Neyman's well-known result on conservative variance estimation to a setting with interference. The additional covariance terms in $R_n$ weighted by $\mathbf{1}\{\ell_{\bm{A}}(i,j)\leq b_n\}$ account for dependence due to interference.

\begin{remark}\label{superpop}
  In the \autoref{sas}, we compare $R_n$ with the bias of the \cite{aronow_estimating_2017} estimator for correctly specified exposure mappings. Simulation results there show that our bias is positive but can be notably smaller than theirs. More generally, the asymptotic behavior of $R_n$ depends on the superpopulation model towards which our framework is agnostic. Since $R_n$ has the form of a network HAC, we expect that it typically converges to the population variance of the unit-level exposure effects, although this requires additional weak dependence conditions on the distributions of $Y_i(\bm{d})$ and $\bm{A}$. Some such conditions are given in Theorem 4.2 of \cite{leung2019inference}.
\end{remark}

To show consistency of $\hat\sigma^2$, define
\begin{equation*}
  \mathcal{J}_n(s, m) = \left\{ (i,j,k,l) \in \mathcal{N}_n^4\colon k\in \mathcal{N}_{\bm{A}}(i, m), l \in \mathcal{N}_{\bm{A}}(j, m), \ell_{\bm{A}}(i,j) = s \right\}.
\end{equation*}

\noindent This is similar to, and evidently contains, $\mathcal{H}_n(s,m)$ from \autoref{psiweak}. 

\begin{assump}[Weak Dependence for $\hat\sigma^2$]\label{psiweak2}
  (a) $\sum_{s=0}^n M_n^\partial(s) \tilde\theta_{n,s}^{1-\epsilon} = O(1)$ for some $\epsilon>0$, (b) $M_n(b_n,1) = o(n^{1/2})$, (c) $M_n(b_n,2) = o(n)$, (d) $\sum_{s=0}^n \abs{\mathcal{J}_n(s, b_n)} \tilde\theta_{n,s} = o(n^2)$.
\end{assump}

\noindent In \autoref{sbandchoice}, we use these conditions to derive \eqref{ourb} and illustrate their plausibility for some general classes of graphs. Part (a) strengthens \autoref{finitevar}, while (b)--(d) regulate $b_n$. Part (b) allows us to replace $\hat\tau(t,t')$ in $\hat\sigma^2$ with its expectation. 
%Parts (a) and (c) are stronger analogs of conditions in Assumptions \ref{finitevar} and \ref{psiweak}.
%Part (c) is similar to the second requirement of \autoref{psiweak} but with $b_n$ in place of $m_n$. 
Part (d) is used to derive the asymptotic bias. It is very similar to the first requirement of \autoref{psiweak}, with $b_n$ and $\mathcal{J}_n(s,\cdot)$ in place of $m_n$ and $\mathcal{H}_n(s,\cdot)$, respectively.

\begin{theorem}[Variance Estimator]\label{bootvar}
  If $b_n \rightarrow \infty$ as $n\rightarrow\infty$, then under Assumptions \ref{aKexposure}--\ref{aani} and \ref{psiweak2}, \eqref{decomp} and \eqref{bootconsist} hold.
\end{theorem}

%----------------------------------------------------------------------
\section{Numerical Illustrations}\label{snumill}
%---------------------------------------------------------------------- 

%---------------------------------------
\subsection{Empirical Application}\label{sapp}
%---------------------------------------

We revisit a network experiment analyzed in \cite{paluck2016changing} and \cite{aronow_estimating_2017} that studies the effect of an anti-conflict intervention on adolescent social norms for antagonistic behavior, including harassment, rumor-mongering, social exclusion, and bullying. In the experimental design, 28 of 56 schools are first randomized into treatment. Then within treated schools, a subset of students are selected as eligible for treatment based on covariates, and half of eligibles are block-randomized into treatment. Treated students are invited to participate in bi-monthly meetings that follow an anti-conflict curriculum designed in part by the researchers of the study. At these meetings, a trained adult leader helps students identify social conflicts at their school and design strategies to reduce conflict. 

\cite{aronow_estimating_2017} and part of the analysis of \cite{paluck2016changing} examine the causal effect of the offer to participate on endorsement of anti-conflict norms. This is measured by self-reports of wearing a wristband disseminated as part of the program as a reward to students observed engaging in conflict-mitigating behavior. Through the course of the experiment, over 2500 wristbands were disseminated and tracked. We study exposure effects similar to those used in the application of \cite{aronow_estimating_2017}. Unlike their analysis, we restrict the data to the five largest treated schools to illustrate what can be learned from data on a few large networks. In each of our schools, the number of eligibles is exactly 64. 

We estimate a treatment and a spillover effect. For the latter, the exposure mapping is $T_i = \ind\{\sum_j A_{ij}D_j > 0\}$, an indicator for whether at least one friend is offered treatment. As in \cite{aronow_estimating_2017}, to ensure overlap, we restrict to the ``spillover population'' consisting of students that have at least one eligible friend ($n=1685$). For the treatment effect, the exposure mapping is $T_i = D_i$, and we restrict to the ``treatment population'' consisting of students eligible for treatment ($n=320$). For both, we compute path distances and average degree in the variance estimator (discussed below) using the full network.

Networks are measured by asking students to name up to ten students at the school ``whom they chose to spend time with in the last few weeks, either in school, out of school, or online.'' Consequently, $\bm{A}$ is directed. When computing the number of treated friends for the exposure mappings, we use the directionality of links. However, when computing network neighborhoods for our variance estimator, we ignore the directionality of links to conservatively define larger neighborhoods and avoid taking a stance on neighborhood definitions for directed networks. 

We next provide some summary statistics. Within the treatment population, the average outcome $Y_i$ is 0.16 (SD 0.37), and by block randomization, exactly 50 percent are treated. Within the spillover population, the average outcome is 0.11 (SD 0.32), and 58 percent (SD 0.49) have at least one treated friend. The data includes the blocks in which eligible students are block-randomized, so we can compute the propensity scores $\pi_i(t)$ for each student using the hypergeometric distribution. For the exposure mapping $T_i = \ind\{\sum_j A_{ij}D_j > 0\}$, given $N$ eligible neighbors, $\pi_i(0)$ is the chance of having 0 out of $N$ successes when drawing without replacement. We find $n^{-1} \sum_{i=1}^n \pi_i(1) = 0.597$, which is close to empirical proportion of 58 percent.

The average out-degree $n^{-1} \sum_{i,j} A_{ij}$ is 7.96. The APL is small, on average 3.37 across our five schools. Since there are $n=3306$ students, $\log n / \log \delta(\bm{A}) = 3.96$, which is very close to 3.37. Thus, given $K=1$, our suggested bandwidth \eqref{ourb} is $b_n=2$. We report results for the range of bandwidths $\{0,\dots,3\}$, noting that $0$ corresponds to standard errors in the no-interference case.
% Note that we're computing the bandwidth based on the APL for the whole network, not the subnetwork on the sample. Think of it as: we're changing the estimand to being conditional on being eligible or conditional on having an eligible neighbor. But we're not changing the overall population actually.

\begin{table}[ht]
\centering
\caption{Estimates and SEs}
\begin{threeparttable}
\begin{tabular}{lrr}
\toprule
{} &  Treatment &  Spillover \\
\cmidrule{2-3}
$\hat\tau(1,0)$ &     0.1500 &     0.0407 \\
$\hat\mu(1)$    &     0.2375 &     0.1293 \\
$\hat\mu(0)$    &     0.0875 &     0.0885 \\
$b_n = 0$       &     0.0443 &     0.0167 \\
$b_n = 1$       &     0.0460 &     0.0184 \\
$b_n = 2$       &     0.0394 &     0.0205 \\
$b_n = 3$       &     0.0470 &     0.0170 \\
\bottomrule
\end{tabular}
\begin{tablenotes}[para,flushleft]
\footnotesize Columns display results for the treatment ($n=320$) and spillover ($n=1685$) effects. Rows ``$b_n=k$'' report SEs for the indicated bandwidths.
\end{tablenotes}
\end{threeparttable}
\label{tempresults}
\end{table}

\autoref{tempresults} presents the results. The first row is the IPW estimator for the indicated exposure effect, and the last four rows are standard errors for the indicated bandwidths. We find a large treatment effect of 0.15, which is significant at the 5 percent level across all bandwidths. The spillover effect is smaller at 0.04, with larger standard errors, and is statistically insignificant for our suggested bandwidth $b_n=2$ at the 5 percent level. The small spillover estimate is largely in line with the estimates implied by Figure 3C of \cite{paluck2016changing}. It does not contradict the overall message of their paper since they find, for example, sizeable spillover effects when comparing treated and untreated schools. In contrast, our analysis so far only makes comparisons within treated schools, using only a subsample of five schools. 

To compare treated and untreated schools, note that for the latter, $Y_i=0$ for all $i$ by design. Then our estimate of $\hat\mu(0)$ for the treatment effect shows that even untreated units are 8.8 percentage points more likely to wear wristbands in treated compared to untreated schools. We obtain standard errors for $\hat\mu(0)$ by replacing $Z_i$ and $\hat\tau(t,t')$ in \eqref{bootstrap} with $\ind_i(t) \pi_i(t)^{-1} Y_i$ and $\hat\mu(0)$. For $b_n = 0,\dots,3$, the standard errors range from $0.012$ to $0.017$, all of which imply a statistically significant effect. Overall, these results indicate that, despite the potential conservativeness of our estimator due to the bias term $R_n$, they can still deliver reasonable standard errors.

%---------------------------------------
\subsection{Monte Carlo}\label{smc}
%---------------------------------------

To study the finite sample properties of our estimators, we simulate data from the two response models studied in \autoref{splaus} using two models of network formation calibrated to the school data from \autoref{sapp}. For the linear-in-means model, $Y_i = V_i(\bm{D},\bm{A},\bm{\varepsilon})$ for
\begin{equation*}
  V_i(\bm{D},\bm{A},\bm{\varepsilon}) = \alpha + \beta \frac{\sum_j A_{ij}Y_j}{\sum_j A_{ij}} + \delta \frac{\sum_j A_{ij}D_j}{\sum_j A_{ij}}  +  D_i\gamma + \varepsilon_i
\end{equation*}

\noindent and $(\alpha,\beta,\delta,\gamma) = (-1,0.8,1,1)$. For the complex contagion model, we set $Y_i = \ind\{V_i(\bm{D},\bm{A},\bm{\varepsilon}) > 0\}$ and $(\alpha,\beta,\delta,\gamma) = (-1,1.5,1,1)$. 

We simulate $\bm{A}$ from configuration and random geometric graph (RGG) models. The former is calibrated to the empirical out-degree sequence $(\sum_{j=1}^n A_{ij})_{i=1}^n$ of the schools used in \autoref{sapp}. This model (approximately) draws an undirected network uniformly at random from the set of all networks with this degree sequence \citep[e.g.][Ch.\ 4.1.4]{jackson2010}. An RGG is a spatial network where units only link with geographically close alters: $A_{ij} = \ind\{\norm{\rho_i-\rho_j} \leq r_n\}$ for $\rho_i \stackrel{iid} \sim \mathcal{U}([0,1]^2)$ and $r_n = (\kappa/(\pi n))^2$. Since $\kappa$ is the limiting expected degree of the model \citep{penrose2003}, we set it equal to the average of the empirical out-degree sequence of the schools. 

Let $\{\nu_i\}_{i=1}^n \stackrel{iid}\sim \mathcal{N}(0,1)$ be independent of $\bm{A}$. For the configuration model, we take $\varepsilon_i = \nu_i$. For the RGG, we instead use $\varepsilon_i = (\rho_{i1}-0.5) + \nu_i$, adding the centered first component of $i$'s ``location'' $\rho_{i1}$. Since units with similar $\rho_{i1}$'s are more likely to form links, this generates unobserved homophily.

Both models yield networks with approximately the same average degree of about 8. For our purposes, the main distinction between them is the APL. The configuration model is theoretically known to have a small APL of logarithmic order in the network size \citep{van2016random}, while the RGG model seems to generate a much larger APL of polynomial order \citep{friedrich2013diameter}. We choose different bandwidths for the two according to \eqref{ourb}. 

Note that the value of the peer effect $\beta$ in both outcome models is actually quite large. The calculations in Appendices \ref{discpsiweak} and \ref{sbandchoice} suggest that it is larger than what our weak dependence conditions actually require under the configuration model (but not the RGG) since it generates a low APL (and hence approximately exponential neighborhood growth rates). Of course, these conditions are likely stronger than necessary, which may explain the good performance in both designs below.

To show different population sizes, we report results using the largest, two largest, and four largest of the treated schools when calibrating the network models. In all cases, we pool the degree sequences across the schools to treat them as one single network. Following the design in \autoref{sapp}, we randomly assign treatments to only units classified as eligible in the data with probability 0.5. 

We compute estimates and standard errors for the spillover effect $\tau(1,0)$ in \autoref{sapp}, whose exposure mapping is $T_i = \ind\{\sum_j A_{ij}D_j > 0\}$, again restricting to the population of units with an eligible neighbor. The standard errors use the bandwidth \eqref{ourb}. Given the APL differences discussed above, which can also be seen in Tables \ref{configtable} and \ref{RGGtable}, the RGG uses $\tilde b_n = \mathcal{L}(\bm{A})^{1/3}$, while the configuration model uses $\tilde b_n = \mathcal{L}(\bm{A})/2$. 

To illustrate the performance of our bandwidth rule across different networks, we redraw $\bm{D}$, $\bm{A}$, and $\bm{\varepsilon}$ for each of the 10k simulation draws. This corresponds to a superpopulation design, so we expect our standard errors to yield confidence intervals with coverage close to the nominal rate of 0.95. We report ``oracle'' standard errors, which correspond to $\var(\hat\tau(1,0))^{1/2}$, approximated by taking the standard deviation of $\hat\tau(1,0)$ over 10k separate simulation draws. We also report ``naive'' i.i.d.\ standard errors to illustrate the degree of dependence in the data.

\begin{table}
\centering
\caption{Simulation Results: Configuration Model}
\begin{threeparttable}
\begin{tabular}{lrrrrrr}
\toprule
{} & \multicolumn{3}{c}{Linear-in-Means} & \multicolumn{3}{c}{Complex Contagion} \\
\cmidrule{2-7}
\# Schools   &               1 &       2 &       4 &               1 &       2 &       4 \\
\midrule
$\hat n(1)$     &         199.3 &  399.7 &  804.2 &         199.3 &  399.7 &  804.2 \\
$\hat n(0)$     &         151.5 &  293.1 &  570.9 &         151.5 &  293.1 &  570.9 \\
$\hat\tau(1,0)$        &           0.311 &    0.317 &    0.313 &           0.077 &    0.077 &    0.080 \\
Our SE          &           1.310 &    0.955 &    0.686 &           0.109 &    0.081 &    0.059 \\
Oracle SE       &           1.402 &    0.986 &    0.694 &           0.117 &    0.084 &    0.060 \\
Our Coverage    &           0.923 &    0.938 &    0.947 &           0.928 &    0.937 &    0.943 \\
Oracle Coverage &           0.947 &    0.950 &    0.954 &           0.948 &    0.950 &    0.951 \\
Naive Coverage  &           0.544 &    0.549 &    0.563 &           0.729 &    0.725 &    0.730 \\
APL             &           3.471 &    3.753 &    4.070 &           3.471 &    3.753 &    4.070 \\
$b_n$    &           2.000 &    2.000 &    2.000 &           2.000 &    2.000 &    2.000 \\
Network Size    &         805 & 1456 & 2725 &         805 & 1456 & 2725 \\
\bottomrule
\end{tabular}
\begin{tablenotes}[para,flushleft]
  \footnotesize Averages over 10k simulations. $\hat n(t) = \sum_i \ind_i(t)$ is the effective sample size of $\hat\mu(t)$. ``Coverage'' rows display empirical coverage for 95\% CIs. ``Naive'' and ``Oracle'' respectively correspond to i.i.d.\ and oracle standard errors.
\end{tablenotes}
\end{threeparttable}
\label{configtable}
\end{table}

Tables \ref{configtable} and \ref{RGGtable} present results for the configuration and RGG models, respectively. ``Our SE'' displays the standard error obtained from our variance estimator and ``Our Coverage'' the empirical coverage of the corresponding two-sided 95\% confidence interval. The effective sample size of $\hat\mu(t)$ is $\hat n(t) = \sum_i \ind_i(t)$. ``Network Size'' is the number of units in the network, whereas the sample size is $\hat n(1) + \hat n(0)$ since, as in the empirical application, we only use units with an eligible neighbor for overlap. 

The results show that the oracle coverage rates are all very close to 0.95, which illustrates the quality of the normal approximation. Our standard errors perform well, with coverage quite close to 0.95. This is in spite of fairly large peer effects $\beta$, which can be seen in the magnitudes of the spillover effect estimates. By contrast, naive standard errors are severely anti-conservative.

\begin{table}
\centering
\caption{Simulation Results: RGG Model}
\begin{threeparttable}
\begin{tabular}{lrrrrrr}
\toprule
{} & \multicolumn{3}{c}{Linear-in-Means} & \multicolumn{3}{c}{Complex Contagion} \\
\cmidrule{2-7}
\# Schools   &               1 &       2 &       4 &               1 &       2 &       4 \\
\midrule
$\hat n(1)$     &         210.0 &  418.2 &  830.1 &         210.0 &  418.2 &  830.1 \\
$\hat n(0)$     &         155.5 &  297.8 &  576.7 &         155.5 &  297.8 &  576.7 \\
$\hat\tau(1,0)$        &           0.702 &    0.700 &    0.700 &           0.088 &    0.089 &    0.089 \\
Our SE          &           1.523 &    1.107 &    0.805 &           0.144 &    0.105 &    0.076 \\
Oracle SE       &           1.601 &    1.142 &    0.824 &           0.152 &    0.108 &    0.079 \\
Our Coverage    &           0.927 &    0.936 &    0.936 &           0.935 &    0.941 &    0.943 \\
Oracle Coverage &           0.953 &    0.950 &    0.947 &           0.950 &    0.951 &    0.954 \\
Naive Coverage  &           0.519 &    0.522 &    0.525 &           0.626 &    0.641 &    0.623 \\
APL             &          13.996 &   18.444 &   24.927 &          13.996 &   18.444 &   24.927 \\
$b_n$    &           2.007 &    3.000 &    3.000 &           2.007 &    3.000 &    3.000 \\
Network Size    &         805 & 1456 & 2725 &         805 & 1456 & 2725 \\
\bottomrule
\end{tabular}
\begin{tablenotes}[para,flushleft]
  \footnotesize See table notes of \autoref{configtable}.
\end{tablenotes}
\end{threeparttable}
\label{RGGtable}
\end{table}

%----------------------------------------------------------------------
\section{Conclusion}\label{sconclude}
%---------------------------------------------------------------------- 

The literature on causal inference under interference typically assumes $K$-neighborhood exposure mappings are correctly specified. This implies a limited model of interference in which units further than distance $K$ from the ego have no effect on the ego's response for some small, known $K$. Such a model is incompatible with those studied in the large theoretical and empirical literature on social interactions, in which units arbitrarily far from the ego can influence the ego's outcome. This paper proposes a richer model of ``approximate neighborhood interference'' (ANI) under which treatments assigned to distant alters have a nonzero effect on the ego's outcome, but the effect declines with network distance. Unlike with existing models of interference, we show that ANI is satisfied by well-known models of social interactions. We also show that ANI is useful for large-sample inference, proving that IPW estimators are consistent and asymptotically normal. 

We consider a network HAC variance estimator for inference robust to misspecification of exposure mappings.  We show the estimator is biased in a finite population setting. The bias term captures the variance of the unit-level treatment effects, which generalizes the well-known result on conservative variance estimation under no interference to settings with dependence due to interference. Finally, we propose a new bandwidth for the HAC estimator, which trades off bias and variance in a manner that adapts to the growth rate of $K$-neighborhoods.  

\appendix
%----------------------------------------------------------------------
\section{Verifying Assumptions}\label{sveri}
%----------------------------------------------------------------------

%---------------------------------------
\subsection{\autoref{psiweak}}\label{discpsiweak}
%---------------------------------------

We verify \autoref{psiweak} for networks with polynomial or exponential neighborhood growth rates. We assume a non-degenerate variance ($\sigma_n^{-2} = O(1)$) and $\theta_{n,s} = e^{-c(1-\epsilon)^{-1} s}$ for some $c> 0$ and $\epsilon\in (0,1)$, which holds for the examples in \autoref{splaus}. We will choose $\epsilon$ in \autoref{psiweak} to be the same as this $\epsilon$.

\paragraph{Polynomial Growth Rate.} Suppose
\begin{equation}
  \sup_n \max_{i\in\mathcal{N}_n} \,\abs{\mathcal{N}_{\bm{A}}(i, s)} = C s^d \label{polynbhd}
\end{equation}

\noindent for $s$ sufficiently large and some $C>0$, $d\geq 1$. Polynomial rates appear to be a property of spatial networks,\footnote{This may require replacing the left-hand side of \eqref{polynbhd} with something weaker such as $\sup_n n^{-1} \sum_{i=1}^n \abs{\mathcal{N}_{\bm{A}}(i, s)}^p$. The max simplifies the arguments in this section, but we expect that more realistic higher-order moment conditions of this type can be used to verify the assumptions.} which are models in which link formation is less likely between spatially distant units. Examples include latent space \citep{hrh2002} and RGG models \citep{penrose2003}. Appendix A of \cite{leung2019inference} shows that, for the RGG, path distance is of the same order as spatial distance for connected units. Since spatial $s$-neighborhoods grow polynomially with $s$, it follows that network $s$-neighborhoods also grow polynomially, with $d$ determined by the spatial dimension.

We next verify \autoref{psiweak} under this setup. Choose $m_n = n^{1/(\alpha d)}$, $\alpha > 4$. First consider the third term in \eqref{3stooges}. This is $O(n^{1.5} e^{-c m_n})$, and hence $o(1)$, since $m_n$ is polynomial in $n$. The second term $n^{-3/2} \sum_{i=1}^n \abs{\mathcal{N}_{\bm{A}}(i, m_n)}^2$ is order $n^{-1/2} m_n^{2d} = n^{2/\alpha - 0.5} = o(1)$. Finally, for the first term in \eqref{3stooges}, observe that we can bound  
\begin{equation}
  \abs{\mathcal{H}_n(s, m_n)} \leq \sum_{i=1}^n \sum_{j\in \mathcal{N}_{\bm{A}}^\partial(i,s)} \abs{\mathcal{N}_{\bm{A}}(i,m_n)} \, \abs{\mathcal{N}_{\bm{A}}(j,m_n)}. \label{HH}
\end{equation}

\noindent This is $O(n\, m_n^{2d} s^d) = O(n^{2/\alpha+1} s^d)$, so the first term in \eqref{3stooges} is $O(n^{2/\alpha-1} \sum_{s=0}^\infty s^d e^{-cs})$.

\paragraph{Exponential Growth Rates.} Suppose
\begin{equation}
  \sup_n \max_{i\in\mathcal{N}_n} \,\abs{\mathcal{N}_{\bm{A}}(i, s)} = C e^{\beta s}
  \label{expnbhd}
\end{equation}

\noindent for $s$ sufficiently large and some $C,\beta > 0$. For $k$-regular tree networks, the left-hand side is exactly $\sum_{m=0}^s k^m$. Inhomogeneous random graphs, a large class of models that includes the Erd\H{o}s-R\'{e}nyi and stochastic block models, have a similar property \citep[see][proofs of Lemmas 14.2 and 14.3]{bollobas_phase_2007}.
% lemma 14.2 shows branching processes have this property. lemma 14.3 shows the network's K-neighborhood sizes are bounded below by those of branching processes so long as the size of the branching process is o(n)
The usual intuition \citep[][Ch.\ 3.8]{barabasi2015} is that the average number of units in a unit's 1-neighborhood is the average degree $\delta(\bm{A}) = n^{-1} \sum_{i,j} A_{ij}$, so the typical number in its $s$-neighborhood is approximately $\delta(\bm{A})^s$, in which case $\beta \approx \log \delta(\bm{A})$ and $C\approx 1$.
%Erd\H{o}s-R\'{e}nyi graphs also have exponential neighborhood growth rates. Roughly the argument is that the expected number of 1-neighbors is order $np$, where $p$ is the linking probability, so the expected number of $s$-neighbors is about order $(np)^s$. In the sparse regime where $p = \lambda/n$ for $\lambda\geq 0$, we have growth rates $\lambda^s$, which are exponential in the limiting expected degree $\lambda$. This can be formalized using a branching process argument, so long as $s = o(n)$ \citep[see e.g.][\S 2.4]{durrett2007random}. Finally, inhomogeneous random graphs, a generalization of Erd\H{o}s-R\'{e}nyi that includes popular stochastic block models, also have exponential growth rates by similar branching process arguments.\footnote{For a formal argument, see the proof of Lemma 14.3 in \cite{bollobas_phase_2007}. On p.\ 93--4, they show that the expected $s$-neighborhood size in an inhomogeneous random graph is bounded below by a branching process that runs for $s$ steps when $s = o(n)$. By the bottom of p.\ 92, this branching process has expected size exponential in $s$.}

We next verify \autoref{psiweak}. Choose $m_n = \alpha\beta^{-1} \log n$, $\alpha \in (1.5\beta c^{-1}, 0.5)$, with $c$ from the definition of $\theta_{n,s}$ above. Such an $\alpha$ exists only if $c>3\beta$, which requires $\theta_{n,s}$ to decay sufficiently fast relative to neighborhood growth rates. The third term in \eqref{3stooges} is order $n^{1.5} e^{-c m_n} = n^{1.5-c\alpha\beta^{-1}} = o(1)$. The second term is order $n^{-1/2} e^{2\beta m_n} = n^{2\alpha - 0.5} = o(1)$. Finally, using \eqref{HH}, the first term of \eqref{3stooges} is conservatively order
\begin{equation*}
  n^{-1} e^{2\beta m_n} \sum_{s=0}^n e^{\beta s} e^{-cs} = n^{2\alpha-1} \sum_{s=0}^n e^{(\beta-c)s} = o(1).
\end{equation*}

%---------------------------------------
\subsection{Choice of Bandwidth}\label{sbandchoice}
%---------------------------------------

We use a mix of formal and heuristic arguments to show that our bandwidth \eqref{ourb} satisfies \autoref{psiweak2}(b)--(d) under different neighborhood growth rates. As in \autoref{discpsiweak}, we suppose that $\sigma_n^{-2} = O(1)$ and $\theta_{n,s} = e^{-c(1-\epsilon)^{-1} s}$ for some $c> 0$ and $\epsilon \in (0,1)$.

\paragraph{Polynomial Growth Rates.} We first consider the case in which $s$-neighborhood sizes grow polynomially with $s$ in the sense of \eqref{polynbhd}. Let $\Delta(\bm{A})$ be the diameter of $\bm{A}$, which is the maximum path length between pairs in the largest component. Then 
\begin{equation}
  \max_i \,\abs{\mathcal{N}_{\bm{A}}(i, \Delta(\bm{A}))} = \alpha n, \label{diamheur}
\end{equation}

\noindent where $\alpha$ is the fraction of units in the largest component of $\bm{A}$. Most real-world networks have a ``giant component,'' meaning $\alpha/n \rightarrow c \in (0,\infty)$ \citep{barabasi2015}. Furthermore, the left-hand side of \eqref{diamheur} is $C\Delta(\bm{A})^d$ under \eqref{polynbhd}. Hence, $\Delta(\bm{A}) = O(n^{1/d})$. This is a well-known heuristic argument for the asymptotic behavior of the diameter or average path length \citep[][Ch.\ 3.8]{barabasi2015}.\footnote{See \cite{friedrich2013diameter} for a formal argument for the RGG.}

Our bandwidth rule \eqref{ourb} is based on the APL $\mathcal{L}(\bm{A})$ rather than the diameter since the former is considered a more robust measurement of network width.\footnote{I thank a referee for this suggestion.} While the heuristics above pertain to the diameter, the derived rate is usually more accurate for the APL. As written in \cite{barabasi2015}, Ch.\ 3.8, ``for most networks [these heuristics offer] a better approximation to the average distance between two randomly chosen nodes, than to [the diameter]. This is because [the diameter] is often dominated by a few extreme paths, while [the APL] is averaged over all node pairs, a process that supresses [sic] the fluctuations.'' Hence, our calculations below take $\mathcal{L}(\bm{A}) \approx n^{1/d}$.

We next verify \autoref{psiweak2}(b)--(d) (the argument for (a) is similar). For (b), using our bandwidth $b_n = \mathcal{L}(\bm{A})^{1/3} \approx n^{1/(3d)}$, $n^{-1} \sum_{i=1}^n \abs{\mathcal{N}_{\bm{A}}(i, b_n)} \approx b_n^d \approx n^{1/3} = o(n^{1/2})$. For (c), $n^{-1}\sum_{i=1}^n \abs{\mathcal{N}_{\bm{A}}(i, b_n)}^2 \approx b_n^{2d} \approx n^{2/3} = o(n)$. For (d), note that \eqref{HH} applies with $\mathcal{J}_n(s,\cdot)$ in place of $\mathcal{H}_n(s,\cdot)$, so
\begin{equation}
  \frac{1}{n^2} \sum_{s=0}^n \abs{\mathcal{J}_n(s, b_n)} \tilde\theta_{n,s} \approx n^{-1} b_n^{2d} \sum_{s=0}^n s^d e^{-c(1-\epsilon)^{-1} s} = O(n^{-1/3}). \label{Jpoly}
\end{equation}

\paragraph{Exponential Growth Rates.} Now suppose \eqref{expnbhd}, and take the typical case of $\beta = \log \delta(\bm{A})$ and $C=1$ discussed after \eqref{expnbhd}. The heuristics following \eqref{diamheur} yield $\Delta(\bm{A}) = \log n / \log \delta(\bm{A})$, as in (3.18) of \cite{barabasi2015}. As discussed above, this heuristic is actually more accurate for the APL, so we take $\mathcal{L}(\bm{A}) \approx \log n / \log \delta(\bm{A})$.\footnote{For a formal argument for inhomogeneous graphs, see Theorem 3.14 of \cite{bollobas_phase_2007}.}

Choosing $b_n$ according to \eqref{ourb} yields $b_n \approx 0.5 \log n / \log \delta(\bm{A})$. Strictly speaking, we will actually need $b_n \approx \alpha \log n / \log \delta(\bm{A})$ for some $\alpha<0.5$, which we will assume next, but since \eqref{ourb} rounds to the nearest integer, there is no difference setting $\alpha=0.5$ in practice. For \autoref{psiweak2}(b), $n^{-1} \sum_{i=1}^n \abs{\mathcal{N}_{\bm{A}}(i, b_n)} \leq e^{b_n \log \delta(\bm{A})} \approx e^{\alpha \log n} = o(n^{1/2})$. \autoref{psiweak2}(c) is similar.
%\frac{1}{n} \sum_{i=1}^n \abs{\mathcal{N}_{\bm{A}}(i, b_n)}^2 \leq e^{2b_n \log \delta(\bm{A})} \approx e^{2\alpha \log n} = o(n).
For (d), if $(1-\epsilon)^{-1}c > \beta = \log \delta(\bm{A})$, which is weaker than the requirement $c>3\beta$ in \autoref{discpsiweak}, then as in \eqref{Jpoly},
\begin{equation*}
  \frac{1}{n^2} \sum_{s=0}^n \abs{\mathcal{J}_n(s, b_n)} \tilde\theta_{n,s} \approx n^{-1} e^{2b_n \log \delta(\bm{A})} \sum_{s=0}^n e^{s \log \delta(\bm{A})} e^{-c(1-\epsilon)^{-1} s} = O(n^{2\alpha-1}).
\end{equation*}

%----------------------------------------------------------------------
\section{AS Variance Estimator}\label{sas}
%----------------------------------------------------------------------

We provide a theoretical comparison of our variance estimator and that of \cite{aronow_estimating_2017} (henceforth AS) and provide simulation evidence on differences in conservativeness. Since their estimator is only valid under correctly specified exposure mappings, we now assume \eqref{reparam}. The estimand is the variance
\begin{align*}
  &\sigma_n^2 = \var(\sqrt{n}\hat\tau(t,t')) = V(t) + V(t') + 2 C(t,t'), \quad\text{where} \\
  &V(t) = \frac{1}{n} \sum_{i=1}^n \tilde Y_i(t)^2 \frac{1-\pi_i(t)}{\pi_i(t)} + \frac{1}{n} \sum_{i=1}^n \sum_{j\neq i} \tilde Y_i(t) \tilde Y_j(t) \frac{\pi_{ij}(t,t)-\pi_i(t)\pi_j(t)}{\pi_i(t)\pi_j(t)}, \\
  &C(t,t') = - \frac{1}{n} \sum_{i=1}^n \sum_{j=1}^n \tilde Y_i(t)\tilde Y_j(t') \frac{\pi_{ij}(t,t')-\pi_i(t)\pi_j(t')}{\pi_i(t)\pi_j(t')},
\end{align*}

\noindent and $\pi_{ij}(t,t') = \E[\ind_i(t)\ind_j(t')]$. The AS estimator given in their equation (11) is
\begin{align*}
  &\hat\sigma^2_{AS} = \hat V(t) + \hat V(t') + 2 \hat C(t,t'), \quad\text{where} \\
  &\begin{aligned} \hat V(t) = \frac{1}{n} \sum_{i=1}^n \frac{Y_i^2 \ind_i(t)}{\pi_i(t)} \frac{1-\pi_i(t)}{\pi_i(t)} &+ \frac{1}{n} \sum_{i=1}^n \sum_{j\neq i} \frac{Y_i \ind_i(t) Y_j \ind_j(t)}{\pi_{ij}(t,t)} \frac{\pi_{ij}(t,t)-\pi_i(t)\pi_j(t)}{\pi_i(t)\pi_j(t)} \ind\{\pi_{ij}(t,t) \neq 0\} \\ &+ \frac{1}{n} \sum_{i=1}^n \sum_{j\neq i} \left( \frac{Y_i^2 \ind_i(t)}{2\pi_i(t)} + \frac{Y_j^2 \ind_j(t)}{2\pi_j(t)} \right) \ind\{\pi_{ij}(t,t)=0\}, \end{aligned} \\
  &\begin{aligned} \hat C(t,t') = - \frac{1}{n} \sum_{i=1}^n \sum_{j\neq i} \frac{Y_i\ind_i(t) Y_j\ind_j(t')}{\pi_{ij}(t,t')} &\frac{\pi_{ij}(t,t')-\pi_i(t)\pi_j(t')}{\pi_i(t)\pi_j(t')} \ind\{\pi_{ij}(t,t')\neq 0\} \\
  &+ \frac{1}{n} \sum_{i=1}^n \sum_{j=1}^n \left( \frac{Y_i^2\ind_i(t)}{2\pi_i(t)} + \frac{Y_j^2\ind_j(t')}{2\pi_j(t')} \right) \ind\{\pi_{ij}(t,t')=0\}. \end{aligned}
\end{align*}

%\noindent The expectation of $\hat\sigma^2_{AS}$ is weakly larger than $\sigma_n^2$ due to the terms involving $\ind\{\pi_{ij}(t,t')=0\}$, which are conservative upper bounds on corresponding terms of $\sigma_n^2$ due to Young's inequality \cite[see e.g.][Proposition 5.2]{aronow_estimating_2017}. Formally, the bias of the AS estimator is
\noindent Noting that $\pi_{ii}(t,t')=0$, $\hat\sigma^2_{AS}$ is conservative for $\sigma_n^2$ with bias
\begin{multline*}
  R_{n,AS} \equiv \E[\hat\sigma^2_{AS}] - \sigma_n^2 = \frac{1}{n} \sum_{i=1}^n (\tilde Y_i(t)-\tilde Y_i(t'))^2 \\
%  &\begin{aligned}+ \frac{1}{n} \sum_{i=1}^n \sum_{j\neq i} &\big( \tilde Y_i(t)\tilde Y_j(t)\ind\{\pi_{ij}(t,t)=0\} - \tilde Y_i(t)\tilde Y_j(t')\ind\{\pi_{ij}(t,t')=0\} \\ & - \tilde Y_i(t')\tilde Y_j(t)\ind\{\pi_{ij}(t',t)=0\} + \tilde Y_i(t')\tilde Y_j(t')\ind\{\pi_{ij}(t',t')=0\} \big) \end{aligned} \nonumber\\
%  &\begin{aligned}+ \frac{1}{n} \sum_{i=1}^n \sum_{j\neq i} &\bigg( \left( \frac{\tilde Y_i(t)^2}{2}+\frac{\tilde Y_j(t)^2}{2} \right) \ind\{\pi_{ij}(t,t)=0\} + \left( \frac{\tilde Y_i(t)^2}{2}+\frac{\tilde Y_j(t')^2}{2} \right)\ind\{\pi_{ij}(t,t')=0\} \\ &+ \left( \frac{\tilde Y_i(t')^2}{2}+\frac{\tilde Y_j(t)^2}{2} \right)\ind\{\pi_{ij}(t',t)=0\} + \left( \frac{\tilde Y_i(t')^2}{2}+\frac{\tilde Y_j(t')^2}{2} \right)\ind\{\pi_{ij}(t',t')=0\}\bigg). \end{aligned} \\
  + 0.5 \frac{1}{n} \sum_{i=1}^n \sum_{j\neq i} \big( (\tilde Y_i(t)+\tilde Y_j(t))^2\ind\{\pi_{ij}(t,t)=0\} + 2(\tilde Y_i(t)-\tilde Y_j(t'))^2 \ind\{\pi_{ij}(t,t')=0\} \\ + (\tilde Y_i(t')+\tilde Y_j(t'))^2 \ind\{\pi_{ij}(t',t')=0\} \big). 
\end{multline*}

\noindent The asymptotic bias of our estimator $\hat\sigma^2$ is $R_n$ given in \eqref{decomp} with $b_n = 2K$ since we are assuming correct specification. It does not appear that $R_n$ and $R_{n,AS}$ can generally be ordered. However, we can consider a few instructive special cases. 

First consider {\bf no interference} ($K=0$). Then $R_n = n^{-1} \sum_{i=1}^n (\tau_i(t,t') - \tau(t,t'))^2$. For the AS estimator, since $\pi_{ij}(t,t') \neq 0$ for all $i\neq j$, $R_{n,AS} = n^{-1} \sum_{i=1}^n \tau_i(t,t')^2$, which is strictly larger than $R_n$ whenever the treatment effect $\tau(t,t')$ is nonzero. Indeed $R_{n,AS}$ is larger by the {\em square} of the treatment effect, so that for large effects and relatively small sample sizes, $\hat\sigma_{AS}^2$ can be substantially more conservative than $\hat\sigma^2$. Next consider {\bf homogeneous unit-level exposure effects}, which corresponds to $\tau_i(t,t') = \tau(t,t')$ for all $i$. In this case, $R_n = 0$, whereas $R_{n,AS}$ is strictly positive if, for example, $\tau_i(t,t') \neq 0$ for all $i$. Finally, consider the {\bf has-treated-neighbor spillover effect} in \autoref{snumill}, where $t=1$, $t'=0$, and $T_i = \ind\{\sum_j A_{ij}D_j > 0\}$. Call a unit $i$ ``eligible'' if $\E[D_i]>0$. Let $\mathcal{E}_{ij}$ be the event that the set of $i$'s eligible neighbors in $\bm{A}$ equals the set of eligible neighbors $i$ has in common with $j$. Then $\pi_{ij}(1,0) = 0$ if and only if $\mathcal{E}_{ij}$ occurs, and $\pi_{ij}(d,d) \neq 0$ for all $d\in\{0,1\}$ and units $i$ and $j$ who have at least one eligible neighbor. Thus, if the population consists of all units with eligible neighbors, as in \autoref{snumill},
\begin{align}
  &R_{n,AS} = \frac{1}{n} \sum_{i=1}^n \tau_i(1,0)^2 + \frac{1}{n} \sum_{i=1}^n \sum_{j\neq i} (\tilde Y_i(1) - \tilde Y_j(0))^2 \ind\{\mathcal{E}_{ij}\}, \quad\text{whereas} \label{ASbias} \\
  &\begin{aligned}R_n = \frac{1}{n} \sum_{i=1}^n &(\tau_i(1,0)-\tau(1,0))^2 \\ &+ \frac{1}{n} \sum_{i=1}^n \sum_{j\neq i} (\tau_i(1,0)-\tau(1,0))(\tau_j(1,0)-\tau(1,0)) \ind\{\ell_{\bm{A}}(i,j) \leq 2\}. \end{aligned} \label{mybias}
\end{align}

\noindent While the first terms of both expressions can be ordered as in the no-interference case, the second terms cannot. As discussed in \autoref{svar}, the second term in \eqref{mybias} accounts for dependence between unit-level exposure effects due to interference, but the second term in \eqref{ASbias} does not have a clear interpretation. However, we can provide simulation evidence on magnitudes.

We simulate networks calibrated to the data in the empirical application, as in \autoref{smc}, but use a different outcome model to impose correct specification: $\tilde Y_i(0) = \varepsilon_i + \sum_j A_{ij}\varepsilon_j / \sum_j A_{ij}$ and $\tilde Y_i(1) = \beta_i + \tilde Y_i(0)$, where $\{\varepsilon_i\}_{i=1}^n \stackrel{iid}\sim \mathcal{N}(0,1)$ is independent of $\{\beta_i\}_{i=1}^n \stackrel{iid}\sim \mathcal{N}(1,1)$. The fraction in $\tilde Y_i(0)$ can be interpreted as exogenous peer effects in unobservables and serves to generate network autocorrelation in baseline outcomes. As in \autoref{smc}, the population is the set of units with eligible neighbors. The left half of \autoref{biastable} reports the average values of $R_{n,AS}$ and $R_n$ across 5000 simulation draws. We see that $R_{n,AS}$ is about 3--5 times larger than $R_n$ on average.

In the previous design, $\tau_i(1,0)=\beta_i$ is independent across $i$. To introduce network autocorrelation in unit-level exposure effects, we instead take $\tilde Y_i(1) = \beta_i + \sum_j A_{ij}\beta_j / \sum_j A_{ij} + \tilde Y_i(0)$. The right half of \autoref{biastable} reports the results. For the configuration model, $R_{n,AS}$ is twice as large as $R_n$ on average, and for the RGG model, it is about 25 percent larger. 

\begin{table}[ht]
\centering
\caption{Comparison of Average Bias}
\begin{threeparttable}
\begin{tabular}{lrrrrrrrr}
\toprule
{} & \multicolumn{4}{c}{Independent Effects} & \multicolumn{4}{c}{Autocorrelated Effects} \\
\cmidrule{2-9}
{} & \multicolumn{2}{c}{Configuration} & \multicolumn{2}{c}{RGG Model} & \multicolumn{2}{c}{Configuration} & \multicolumn{2}{c}{RGG Model} \\
\cmidrule{2-9}
\# Schools   &               1 &       2 &      1 &       2 &          1 &       2 &      1 &       2 \\
\midrule
$R_{n,AS}$  &          9.4 & 10.2 & 7.1 & 7.6 &          154.6 & 158.7 & 170.0 & 180.0 \\
$R_n$       &          1.9 & 1.9 & 2.0 & 2.0 &          67.6 & 73.8 &  136.6 & 144.1 \\
$n$         &          350.8 & 692.8 & 365.5 & 716.1 &          350.8 & 692.8 & 365.5 & 716.1 \\
\bottomrule
\end{tabular}
\begin{tablenotes}[para,flushleft]
  \footnotesize Cells are averages over 5k simulations. $n=$ \# units with eligible neighbors.
\end{tablenotes}
\end{threeparttable}
\label{biastable}
\end{table}

%----------------------------------------------------------------------
\section{Proofs}\label{sproofs}
%----------------------------------------------------------------------

\begin{proof}[Proof of \autoref{linearinmeans}]
  The reduced form of the linear-in-means model is
  \begin{multline*}
    Y_i(\bm{D}) = \alpha + D_i\gamma + \varepsilon_i + \ind\left\{\sum_j A_{ij} > 0\right\} \times \\
    \left( \frac{\alpha\beta}{1-\beta} + \gamma\beta \sum_{k=0}^\infty \beta^k \left( \sum_{j_1=1}^n \frac{A_{ij_1}}{\sum_\ell A_{i\ell}} \sum_{j_2=1}^n \frac{A_{j_1j_2}}{\sum_\ell A_{j_1\ell}} \cdots \sum_{j_{k+1}=1}^n \frac{A_{j_kj_{k+1}}}{\sum_\ell A_{j_k\ell}} D_{j_{k+1}} \right) \right. \\
    \left. + \sum_{k=1}^\infty \beta^k \left( \sum_{j_1=1}^n \frac{A_{ij_1}}{\sum_\ell A_{i\ell}} \sum_{j_2=1}^n \frac{A_{j_1j_2}}{\sum_\ell A_{j_1\ell}} \cdots \sum_{j_k=1}^n \frac{A_{j_{k-1}j_k}}{\sum_\ell A_{j_{k-1}\ell}} \varepsilon_{j_k} \right) \right) 
  \end{multline*}

  \noindent for $j_0=i$. Consider a counterfactual linear-in-means model in which the set of units is $\mathcal{N}_{\bm{A}}(i,s)$ rather than $\mathcal{N}_n$. That is, outcomes are realized according to \eqref{LIMM} but with primitives $(\bm{D}_{\mathcal{N}_{\bm{A}}(i,s)}, \bm{A}_{\mathcal{N}_{\bm{A}}(i,s)}, \bm{\varepsilon}_{\mathcal{N}_{\bm{A}}(i,s)})$ rather than $(\bm{D},\bm{A},\bm{\varepsilon})$, where $\bm{\varepsilon} = (\varepsilon_i)_{i=1}^n$ and $\bm{\varepsilon}_{\mathcal{N}_{\bm{A}}(i,s)} = (\varepsilon_i\colon i\in \mathcal{N}_{\bm{A}}(i,s))$. Let $Y_i^{(s)}(\bm{D})$ be unit $i$'s outcome in the counterfactual model. To compare $Y_i^{(s)}(\bm{D})$ with $Y_i(\bm{D})$, consider the $k$th term in the first series of the previous equation (the term multiplying $\gamma\beta \cdot \beta^k$). We can rewrite this as $\sum_{j=1}^n \omega_{ij}^k D_j$, where $\omega_{ij}^k$ is the following weighted sum of all walks of length $k+1$ from $i$ to $j$:
  \begin{equation*}
    \omega_{ij}^k = \sum_{j_1=1}^n \sum_{j_2=1}^n \cdots \sum_{j_k=1}^n \frac{A_{ij_1}}{\sum_\ell A_{i\ell}} \frac{A_{j_1j_2}}{\sum_\ell A_{j_1\ell}} \cdots \frac{A_{j_kj}}{\sum_\ell A_{j\ell}}.
  \end{equation*}

  \noindent In going from $Y_i(\bm{D})$ to $Y_i^{(s)}(\bm{D})$, we lose terms in $\omega_{ij}^k$ involving walks that traverse paths of length greater than $s$. We can conservatively bound this loss by including {\em all} walks with length exceeding $s$ in the loss. Then since $\sum_{j=1}^n \omega_{ij}^k = 1$ 
  % follows from \tilde\bm{A}^k 1 = 1
  and $D_j$ is binary,
  \begin{equation*}
    \abs{Y_i(\bm{D}) - Y_i^{(s)}(\bm{D})} \leq \gamma\beta \sum_{k=s+1}^\infty \abs{\beta}^k + \sup_i\, \abs{\varepsilon_i} \sum_{k=s+1}^\infty \abs{\beta}^k
  \end{equation*}

  \noindent for $s\geq 1$. Since $\varepsilon_i$ is uniformly bounded, the right-hand side is bounded by a constant times $\abs{\beta}^s$. Furthermore, the argument above is the same if we replace $\bm{D}$ with $\bm{D}^{(i,s)}$, the latter defined prior to \autoref{aani}. Since $Y_i^{(s)}(\bm{D}) = Y_i^{(s)}(\bm{D}^{(i,s)})$, the result follows from the triangle inequality.
\end{proof}

%---------------------
\begin{proof}[Proof of \autoref{thresholdmodel}]
  The arguments that follow borrow ideas from the proofs of Proposition 1 of \cite{xl2015} and Theorem 6.1 of \cite{leung2019inference}. Let $\tilde{\bm{G}}$ be the directed network with $ij$ entry $A_{ij}\varphi_j$, $C_i$ the set of units in the strongly connected component of $\tilde{\bm{G}}$ containing $i$, and
  \begin{equation*}
    C_i^+ = C_i \medcup \left\{ k \in \mathcal{N}_n\colon \exists j \in C_i \text{ such that } A_{jk}(1-\varphi_i)=1  \right\}.
  \end{equation*}

  \noindent Let $Y_i(\bm{D})$ be $i$'s outcome under the complex contagion model with $n$ units. As in the proof of our \autoref{linearinmeans}, consider a counterfactual model in which the set of units is $\mathcal{N}_{\bm{A}}(i,s)$ instead of $\mathcal{N}_n$, meaning outcomes are realized according to the same model but with primitives $(\bm{D}_{\mathcal{N}_{\bm{A}}(i,s)}, \bm{A}_{\mathcal{N}_{\bm{A}}(i,s)},$ $\bm{\varepsilon}_{\mathcal{N}_{\bm{A}}(i,s)})$ instead of $(\bm{D},\bm{A},\bm{\varepsilon})$. Let $Y_i^{(s)}(\bm{D})$ be $i$'s outcome in this counterfactual model. Key to our argument is the fact that 
  \begin{equation}
    Y_i(\bm{D}) = Y_i^{(s)}(\bm{D}) \quad\text{if}\quad \bm{A}_{C_i^+} \subseteq \bm{A}_{\mathcal{N}_{\bm{A}}(i,s)}, \label{stratneigh}
  \end{equation}

  \noindent where $\bm{A}_{C_i^+} \subseteq \bm{A}_{\mathcal{N}_{\bm{A}}(i,s)}$ means the former is a subnetwork of the latter. 
  
  To show \eqref{stratneigh}, let $Y_i^t(\bm{D})$ be $i$'s outcome at time $t$ of the dynamic process described prior to the proposition for the model with $n$ units. If $\varphi_i=0$, then $i$ has a dominant strategy $Y_i^*(\bm{D})$, so $Y_i^t(\bm{D}) = Y_i^*(\bm{D})$ for all $t>0$. If instead $\varphi_i=1$, then $i$'s outcome may potentially change at any period $t>0$ in the process, depending on the outcomes of neighboring units at $t-1$. Consider any path in $\bm{A}$ connecting units $i$ and $j$. If $\varphi_k=1$ for all units $k$ along that path, then unit $i$'s outcome may change at any period $t$ in the dynamic process, depending on the outcome of $j$ at some prior period. However, if for all such paths, there exists some unit $k$ along that path such that $\varphi_k=0$, then unit $i$'s outcome will never be affected by unit's $j$ outcome at any past period. Now, if $j \not\in C_i^+$, then by construction, there exists such a unit $k$ along any path connecting $i$ and $j$. Therefore, $Y_i(\bm{D})$ is invariant to the removal of units $\mathcal{N}_n\backslash C_i^+$ from the model in the sense of \eqref{stratneigh}. It follows that
  \begin{multline}
    \E[\abs{Y_i(\bm{D}) - Y_i^{(s)}(\bm{D})}] \leq \prob(Y_i(\bm{D}) \neq Y_i^{(s)}(\bm{D})) \leq \prob(\bm{A}_{C_i^+} \not\subseteq \bm{A}_{\mathcal{N}_{\bm{A}}(i,s)}) \\
    \leq \sum_{j_1 \neq \dots \neq j_{s-1}} A_{ij_1} \E[\varphi_{j_1}] A_{j_1j_2} \E[\varphi_{j_2}] \cdot \dots \cdot A_{j_{s-3}j_{s-1}} \E[\varphi_{j_{s-1}}] = \sum_j (\bm{G}^{s-1})_{ij}. \label{TSIinequal}
  \end{multline}

  \noindent The third inequality follows from the union bound, independence of treatments, and the fact that $\bm{A}_{C_i^+} \not\subseteq \bm{A}_{\mathcal{N}_{\bm{A}}(i,s)}$ implies there exists a path of length at least $s-1$ starting from $i$ such that $\varphi_k=1$ for all units $k$ on that path. % s-1 because it could be that the only node in C_i^+ outside the s-nbhd is one with a robust action (for whom \varphi_k = 0)

  For all $s$, $\abs{Y_i(\bm{D}) - Y_i^{(s)}(\bm{D})} \leq 1$. For $s-1\geq \bar{s}$, $\eqref{TSIinequal} \leq \rho_n(\bar{s})^{s-1}$. Furthermore, these arguments hold if we replace $\bm{D}$ with $\bm{D}^{(i,s)}$, the latter defined prior to \autoref{aani}. Since $Y_i^{(s)}(\bm{D}) = Y_i^{(s)}(\bm{D}^{(i,s)})$, the result follows from the triangle inequality.
\end{proof}

%---------------------
\begin{proof}[Proof of \autoref{anipsi}]
  Let either $h,h' \in \N$, $f \in \mathcal{L}_h$, and $f' \in \mathcal{L}_{h'}$, or $h=h'=1$ and $f=f'=i^*$. Let $s>0$ and $(H,H') \in \mathcal{P}_n(h,h';s)$. Define $\xi = f(\bm{Z}_H)$ and $\zeta = f'(\bm{Z}_{H'})$. Let $\bm{D}', \bm{D}''$ each be independent copies of $\bm{D}$. Define $\bm{D}^{(i,s,\xi)} = (\bm{D}_{\mathcal{N}_{\bm{A}}(i,s)}, \bm{D}_{-\mathcal{N}_{\bm{A}}(i,s)}')$, $\bm{D}^{(i,s,\zeta)} = (\bm{D}_{\mathcal{N}_{\bm{A}}(i,s)}, \bm{D}_{-\mathcal{N}_{\bm{A}}(i,s)}'')$, and 
  \begin{align*}
    &Z_i^{(s,\xi)} = Y_i(\bm{D}^{(i,s,\xi)}) \left( \frac{\ind\{T(i,\bm{D}^{(i,s,\xi)},\bm{A})=t\}}{\pi_i(t)} - \frac{\ind\{T(i,\bm{D}^{(i,s,\xi)},\bm{A})=t'\}}{\pi_i(t')} \right), \\
    &Z_i^{(s,\zeta)} = Y_i(\bm{D}^{(i,s,\zeta)}) \left( \frac{\ind\{T(i,\bm{D}^{(i,s,\zeta)},\bm{A})=t\}}{\pi_i(t)} - \frac{\ind\{T(i,\bm{D}^{(i,s,\zeta)},\bm{A})=t'\}}{\pi_i(t')} \right).
  \end{align*}

  \noindent Finally, let $\xi^{(s)} = f( (Z_i^{(s,\xi)} \colon i \in H) )$ and $\zeta^{(s)} = f'( (Z_i^{(s,\zeta)} \colon i \in H') )$. 
  
  Since $Z_i$ is uniformly bounded by Assumptions \ref{aoverlap} and \ref{aYbound}, $\abs{\cov(\xi,\zeta)} \leq 2\norm{f}_\infty \norm{f'}_\infty$, so for $s \leq 2\max\{K,1\}$, we have $\abs{\cov(\xi,\zeta)} \leq \psi_{h,h'}(f,f')$. Now consider $s>2\max\{K,1\}$, so that $\ell_{\bm{A}}(H,H') > 2\max\{K,1\}$. By \autoref{aKexposure}, $(Z_i^{(\lfloor s/2 \rfloor,\xi)} \colon i \in H) \indep (Z_j^{(\lfloor s/2 \rfloor,\zeta)} \colon j \in H')$. Then 
  \begin{align*}
    \abs{\cov(\xi,\zeta)} &\leq \abs{\cov(\xi-\xi^{(\lfloor s/2 \rfloor)}, \zeta)} + \abs{\cov(\xi^{(\lfloor s/2 \rfloor)}, \zeta-\zeta^{(\lfloor s/2 \rfloor)})} \\ % first add/subtract \xi^{(\lfloor s/2 \rfloor)} and split into two. then in the second, add/subtract \zeta^{(\lfloor s/2 \rfloor)} and split into two. second term is cov(\xi^{(\lfloor s/2 \rfloor)},\zeta^{(\lfloor s/2 \rfloor)})=0
			  &\leq 2 \norm{f'}_\infty \E[\abs{\xi-\xi^{(\lfloor s/2 \rfloor)}}] + 2 \norm{f}_\infty \E[\abs{\zeta-\zeta^{(\lfloor s/2 \rfloor)}}] \\
			  &\leq 2 \left(h \norm{f'}_\infty \text{Lip}(f) + h' \norm{f}_\infty \text{Lip}(f')\right) \theta_{n,\lfloor s/2 \rfloor}.
  \end{align*}

  \noindent The last line uses the fact that, by \autoref{aKexposure},
  \begin{equation*}
    \frac{\ind\{T(i,\bm{D}^{(i,\lfloor s/2 \rfloor,\xi)},\bm{A})=t\}}{\pi_i(t)} = \frac{\ind\{T(i,\bm{D}^{(i,\lfloor s/2 \rfloor,\zeta)},\bm{A})=t\}}{\pi_i(t)} = \frac{\ind\{T(i,\bm{D},\bm{A})=t\}}{\pi_i(t)},
  \end{equation*}
  
  \noindent for any $t\in\mathcal{T}$, and by \autoref{aani}, $\max_{i\in\mathcal{N}_n} \E[\abs{Y_i(\bm{D}) - Y_i(\bm{D}^{(i,\lfloor s/2 \rfloor,\xi)})}] \leq \theta_{n,\lfloor s/2 \rfloor}$.
\end{proof}

%---------------------
\begin{proof}[Proof of \autoref{lln}]
  Since $\E[\hat\tau(t,t')] = \tau(t,t')$, we only need to show that $\var(\hat\tau(t,t')) = o(1)$. Since treatments are independent across units, by \autoref{aani}, $\cov(Z_i,Z_j)=0$ if $\ell_{\bm{A}}(i,j)>n-1$. Hence,
  \begin{equation*}
    \var(\hat\tau(t,t')) = \frac{1}{n^2} \sum_{i=1}^n \var(Z_i) + \sum_{s=1}^{n-1} \frac{1}{n^2} \sum_{i=1}^n \sum_{j\neq i} \ind\{\ell_{\bm{A}}(i,j)=s\} \cov(Z_i,Z_j).
  \end{equation*}

  \noindent Using \autoref{anipsi} and uniform boundedness of $Z_i$ (Assumptions \ref{aoverlap} and \ref{aYbound}), the right-hand side is bounded above by $C(n^{-1} + n^{-2}\sum_{s=1}^{n-1} \tilde\theta_{n,s} \sum_{i=1}^n \abs{\mathcal{N}^\partial_{\bm{A}}(i,s)})$ for some universal $C>0$, and this is $o(1)$ by \autoref{finitevar}. 
\end{proof}

%---------------------
\begin{proof}[Proof of \autoref{clt}]
  Apply our \autoref{anipsi} and Theorem 3.2 of \cite{kojevnikov2019limit}.
\end{proof}

%---------------------
\begin{proof}[Proof of \autoref{bootvar}]
  Observe that \eqref{bootconsist} follows from Proposition 4.1 of \cite{kojevnikov2019limit} since our \autoref{psiweak2} implies their Assumption 4.1. To establish \eqref{decomp}, note that there are two parts of alleged $o_p(1)$ term in \eqref{decomp}. The first is due to replacing $\hat\tau(t,t')$ with $\tau(t,t')$ in the formula for $\hat\sigma^2$. This replacement creates a remainder term of the form
  \begin{multline*}
    r_n = \frac{2}{n} \sum_{i=1}^n \sum_{j=1}^n (Z_i - \tau(t,t')) (\tau(t,t') - \hat\tau(t,t')) \mathbf{1}\{\ell_{\bm{A}}(i,j)\leq b_n\} \\ + (\tau(t,t') - \hat\tau(t,t'))^2 \frac{1}{n} \sum_{i=1}^n \sum_{j=1}^n \mathbf{1}\{\ell_{\bm{A}}(i,j)\leq b_n\}.
  \end{multline*}
  
  \noindent Since $Z_i$ is uniformly bounded (Assumptions \ref{aoverlap} and \ref{aYbound}), for some $C>0$ and any $n$,
  \begin{equation*}
    \abs{r_n} \leq C \abs{\tau(t,t') - \hat\tau(t,t')}\, \frac{1}{n} \sum_{i=1}^n \sum_{j=1}^n \mathbf{1}\{\ell_{\bm{A}}(i,j)\leq b_n\}.
  \end{equation*}

  \noindent The summation term is equal to $M_n(b_n,1)$. The term in the absolute value is $O_p(n^{-1/2})$ since \autoref{psiweak2}(a) implies that $\var(\hat\tau(t,t')) = O(n^{-1})$ (see the proof of \autoref{lln}). Hence, the previous display is $o_p(1)$ by \autoref{psiweak2}(b).

  The remaining parts of the alleged $o_p(1)$ term in \eqref{decomp} are the cross-terms 
  \begin{equation*}
    \frac{2}{n} \sum_{i=1}^n \sum_{j=1}^n (Z_i-\tau_i(t,t'))(\tau_j(t,t')-\tau(t,t')) \mathbf{1}\{\ell_{\bm{A}}(i,j)\leq b_n\}.
  \end{equation*}

  \noindent We show this is $o_p(1)$. For $W_i = \sum_{j=1}^n (\tau_j(t,t')-\tau(t,t')) \mathbf{1}\{\ell_{\bm{A}}(i,j)\leq b_n\}$, 
  \begin{multline*}
    \E\bigg[ \bigg| \frac{1}{n} \sum_{i=1}^n \sum_{j=1}^n (Z_i-\tau_i(t,t'))(\tau_j(t,t')-\tau(t,t')) \mathbf{1}\{\ell_{\bm{A}}(i,j)\leq b_n\} \bigg| \bigg] \\
    \leq \E\bigg[ \bigg( \frac{1}{n} \sum_{i=1}^n (Z_i-\tau_i(t,t'))W_i \bigg)^2 \bigg]^{1/2} \\
    \leq \bigg( \frac{1}{n^2} \sum_{i=1}^n \var(Z_i)W_i^2 + C \frac{1}{n^2} \sum_{s=0}^n \tilde\theta_{n,s} \sum_{i=1}^n \sum_{j\neq i} \ind\{\ell_{\bm{A}}(i,j)=s\} \abs{W_i W_j} \bigg)^{1/2}
  \end{multline*}

  \noindent for some $C>0$ by \autoref{anipsi}. Since $Z_i$ is uniformly bounded, for some $C'>0$, $n^{-2} \sum_{i=1}^n \var(Z_i)W_i^2 \leq C'n^{-1} M_n(b_n,2)$, which is $o(1)$ by \autoref{psiweak2}(c). Likewise,
  \begin{equation*}
    \frac{1}{n^2} \sum_{s=0}^n \tilde\theta_{n,s} \sum_{i=1}^n \sum_{j\neq i} \ind\{\ell_{\bm{A}}(i,j)=s\} \abs{W_i W_j} \leq \frac{C''}{n^2} \sum_{s=0}^n \tilde\theta_{n,s} \mathcal{J}_n(s,b_n)
  \end{equation*}

  \noindent for some $C''>0$, and this is $o(1)$ by \autoref{psiweak2}(d).
\end{proof}

%----------------------------------------------------------------------

\FloatBarrier
\phantomsection
\addcontentsline{toc}{section}{References}
\bibliography{approx_nbhd}{} 
\bibliographystyle{aer}

%----------------------------------------------------------------------

\end{document}